\documentclass[a4paper,11pt]{article}
\usepackage{graphicx}
\usepackage{mathtools}
\usepackage{jheppub}
\usepackage[T1]{fontenc}
\usepackage{braket}	
\usepackage{dcolumn}
\usepackage{comment}

\newcommand{\be}{\begin{equation}}
\newcommand{\ee}{\end{equation}}

\newcommand{\lt}{\left}
\newcommand{\rt}{\right}
\newcommand{\del}{\partial}

\newcommand{\non}{\nonumber \\}

\newcommand{\fn}{\footnote}

\title{Low energy limit from high energy expansion in mass gapped theory}

\author{Hiromasa Takaura}

\affiliation[1]{Center for Gravitational Physics and Quantum Information, Yukawa Institute for Theoretical Physics, Kyoto University, Kyoto 606-8502, Japan}

\emailAdd{hiromasa.takaura@yukawa.kyoto-u.ac.jp}

\abstract{
We present a method to extract the low energy behavior of physical observables from their high energy expansions, 
systematically calculable via the operator product expansion (OPE), in asymptotically free and mass-gapped theories. 
By applying the inverse Laplace transform to correlation functions, their analytic structure is modified such that
low-energy information connects with high energy expansions.
Furthermore, this transformation alleviates the renormalon problem,
enabling a more straightforward application of the OPE compared to the OPE before the transformation.
We demonstrate that the low energy limit of correlation functions can be accurately extracted using the OPE
in the two dimensional $O(N)$ nonlinear $\sigma$ model, serving as a first testing ground.
}

\begin{document}

\preprint{YITP-24-42}

\maketitle
\flushbottom

\section{Introduction and key idea}

Asymptotically free theories, including Quantum Chromodynamics (QCD), 
exhibit many nontrivial and interesting aspects at low energies, 
such as color confinement and the mass gap.
However, calculating low energy behaviors analytically in such theories remains a challenge, with
the  exception of a few specially solvable cases.
In QCD, while lattice simulations offer a method for calculating low energy behaviors,
they are inherently numerical and leave a gap for the desire of analytic approaches,
which would provide deeper theoretical insights.

High energy expansion in asymptotically free theories, in contrast, can systematically be calculated,
utilizing the operator product expansion (OPE).
The OPE approximates an observable by the double expansion in the coupling $g(p)$ and $\Lambda/p$,
which are small at high energies $p \gg \Lambda$,  where $p$ is the typical energy scale of an observable and $\Lambda$ is the dynamical scale (which is denoted by $m$ in the main body of this paper). 
However, such high energy expansion hardly offers insights into low energy behaviors $p \ll \Lambda$ even if
it was calculated to the all orders. 
This is because the convergence radius of the OPE is considered to be zero.
This can be inferred from the fact that particle production can occur at an arbitrary high energy, and then
there exist singularities arbitrarily close to the origin of the $1/p^2$-plane.


In this paper, we propose a new analysis method which can explore low energy behaviors from high energy expansions.
The idea is as follows.
Consider a correlation function $D(p^2)$ in an asymptotically free and mass-gapped theory
(e.g. the Adler function in QCD).
Because the closest singularity to the origin in the $p^2$-plane is located at $p^2=-m^2_{\rm gap}$, 
it has a Taylor expansion around the low energy limit $p^2=0$,
\be
D(p^2)=\sum_{n=0}^{\infty} c_n \lt( \frac{p^2}{\Lambda^2} \rt)^n . \label{lowenergyexpansion}
\ee
which converges for $|p^2| < m^2_{\rm gap}$,
where $m_{\rm gap}$ represents the minimal threshold energy in this correlation function.
(In this paper, we employ the Euclidean metric. Therefore negative $p^2$ corresponds to the time-like region.)
In this paper, we consider the inverse Laplace transform with respect to $r \equiv1/p^2$,
\be
\tilde{D}(\tau) \equiv \frac{1}{2 \pi i} \int_{-i \infty}^{i \infty} \frac{dr}{r} D(p^2=1/r) e^{\tau r} . \label{dualtr}
\ee
Now $\tilde{D}(\tau)$ has the small-$\tau$ expansion
\be
\tilde{D}(\tau)=\sum_{n=0}^{\infty} \frac{c_n}{n!} \lt( \frac{\tau}{\Lambda^2} \rt)^n . \label{introBorel}
\ee
Due to $\sim n!$ in the denominator, the convergence radius of this series expansion is infinity,
so that $\tilde{D}(\tau)$ can be defined as an analytic function.
Therefore, the coefficients $c_n$ can be extracted by
\be
c_n=\frac{n!}{2 \pi i} \Lambda^{2n} \oint_{\tau=\tau_0 e^{i \theta}} \frac{d \tau}{\tau^{n+1}} \tilde{D}(\tau) ,
\ee
where the contour integral on a circle in the complex $\tau$-plane, $\tau=\tau_0 e^{i \theta}$ with $-\pi < \theta \leq \pi$ and $\tau_0>0$, is considered.
The remarkable point is that the radius of the contour, $\tau_0$, can be taken arbitrary because $\tilde{D}(\tau)$ is analytic at any point.
Therefore, by taking $\tau_0$ so large that the OPE is valid, 
one could extract $c_n$ by
\be
c_n\overset{?}{=}\frac{n!}{2 \pi i} \Lambda^{2n} \oint_{\tau=\tau_0 e^{i \theta}} \frac{d \tau}{\tau^{n+1}} \tilde{D}_{\rm OPE}(\tau) . \label{naiveexpect}
\ee
It is straightforward to calculate the OPE for $\tilde{D}(\tau)$ based on the definition~\eqref{dualtr}
if the OPE of $D(p^2)$ is available.
This is the key idea to extract the low energy information from the OPE.
(It will turn out that eq.~\eqref{naiveexpect} requires slight modification due to duality violations explained below.)

In this paper, we employ the two-dimensional $O(N)$ nonlinear $\sigma$ model \cite{Bardeen:1976zh} as a first testing ground for the idea
represented by eq.~\eqref{naiveexpect}.
This model shares crucial features with QCD such as asymptotic freedom and mass gap, and 
is solvable in the large-$N$ limit. 
Hence, we can study various aspects of the present method explicitly 
and evaluate the effectiveness of the method.
We study two examples in this model. One is a correlation function where renormalons
do not appear in each step of calculations, 
and the other has renormalons in its Wilson coefficients in the OPE,
like in the case of QCD.
On the other hand, there is a considerable difference from QCD that
the convergence radius of the OPE is non-zero in the large-$N$ limit of the model.
We will discuss later which part of the analyses in the present paper would be changed 
when the method is applied to QCD.


Studying the nonlinear $\sigma$-model, we find that there are serious duality violations (DV)
in the OPE for the inverse Laplace transformed correlators.
While $\tilde{D}_{\rm OPE}(\tau)$ gives a good approximation of the exact result for ${\rm Re} \, [\tau]>0$  as long as $|\tau| \gg \Lambda^2$,
it significantly deviates for $\tau < 0$ even with $|\tau| \gg \Lambda^2$.
We call this deviation DV.
Considering this fact, we need to employ a different equation from eq.~\eqref{naiveexpect} 
for estimate of the low energy expansion coefficients.
We use, e.g.,
\be
c_0 \approx \frac{1}{2 \pi i} \oint_{\tau=\tau_0 e^{i \theta}} \frac{d \tau}{\tau} \lt(\frac{\tau}{\tau_0}+1 \rt)^k \tilde{D}_{\rm OPE}(\tau)  \label{c0modified}
\ee
with a positive integer $k$, where the region of  ${\rm Re} \, [\tau]<0$ is suppressed.
It is demonstrated that the above modified formula indeed gives accurate estimates 
once $k$ is chosen optimally.

Furthermore, one of the remarkable virtues of the inverse Laplace transform is
that the quantity after the transform, $\tilde{D}(\tau)$, gets largely free from 
the so-called renormalon problem.
Actually, the inverse Laplace transform~\eqref{dualtr} has been originally introduced in ref.~\cite{Hayashi:2023fgl}
in order to solve the renormalon problem.
The OPE of $D(p^2)$ is given by
\be
D_{\rm OPE}(p^2)=\sum_{n=0}^{\infty} C_n(p^2) \frac{\langle \mathcal{O}_n \rangle}{(p^2)^{d_n/2}} , \label{OPE}
\ee
where $\langle \mathcal{O}_n \rangle$ is the nonperturbative matrix element of a local operator whose mass dimension is $d_n$
and the Wilson coefficient $C_n(p^2)$ is given by a perturbative series $\sum_{n=0}^{\infty} a_n g^{n+1}(p^2)$.
The perturbative series are generally divergent because of the factorial growth of the perturbative coefficient $a_n \sim b_0^n n!$,
which is called a renormalon divergence. ($b_0$ is the first coefficient of the beta function.)
The renormalon problem indicates that the perturabtive calculations of the Wilson coefficients have 
inevitable uncertainties of $\sim (\Lambda^2/p^2)^i$ with an integer $i$ \cite{Beneke:1998ui}.
In order to obtain an accurate OPE result beyond the renormalon uncertainties, 
one needs to adopt a method beyond fixed order perturbation theory to calculate the Wilson coefficients.
In addition, one needs to determine the matrix elements $\langle \mathcal{O}_n \rangle$,
which are potentially ambiguous and whose uncertainties are considered to be responsible
for the cancellation of the renormalon uncertainties (see, e.g., Refs.~\cite{David:1982qv,Beneke:1998eq,Schubring:2021hrw}), using nonperturbative data.
See refs.~\cite{Ayala:2019uaw,Benitez-Rathgeb:2022yqb,Beneke:2023wkq} 
for recent developments in this direction in QCD. 
However, in the OPE for $\tilde{D}(\tau)$,
\be
\tilde{D}_{\rm OPE}(\tau)=\sum_{n=0}^{\infty} \tilde{C}_n(\tau) \frac{\langle \mathcal{O}_n \rangle}{(\tau^2)^{d_n/2}}  \label{OPEtilde}
\ee
it was shown in ref.~\cite{Hayashi:2023fgl} that
(i) the Wilson coefficients $\tilde{C}_n(\tau)$ do not have renormalons,
and correspondingly, (ii) the ambiguous nonperturbative matrix elements disappear in the OPE~\eqref{OPEtilde},
as long as the renormalon uncertainties are proportional to $\propto (\Lambda^2/p^2)^i$ with an positive integer $i$.
Even when the renormalon uncertainties are given by the form of $\propto g(p^2)^{\nu} (1+\mathcal{O}(g(p^2))) (\Lambda^2/p^2)^i$, 
the renormalon uncertainties and the contribution from the ambiguous matrix elements are suppressed. 
Therefore, an accurate result for $\tilde{D}_{\rm OPE}(\tau)$ can be obtained more straightforwardly compared to $D_{\rm OPE}(p^2)$.
This characteristic enhances the effectiveness of the method, as demonstrated below.

We give remarks on the relation to previous works.
First, we note that the present work is independent from 
the so-called sum rule analyses \cite{Shifman:1978bx},
although some equations look apparently similar.
In sum rule analyses, the Borel transform of {\it{high}} energy expansions of 
observables are considered, while we consider the Borel transform of 
{\it{low}} energy expansions
as in eq.~\eqref{introBorel}.
These two expansions have very different features
in terms of the encoded information and the convergence radius.
Related to this difference, sum rule analyses mainly study the vacuum condensates appearing
as part of the coefficients of the high energy expansions,
while we explore the low energy behavior of the observables themselves.

Secondly, the discussion on renormalons presented in this paper 
was already established in Refs.~\cite{Beneke:1998eq,Hayashi:2023fgl},
and this paper does not aim to provide new insights into renormalons.
The main and new focus of this paper is the application of the inverse Laplace transform
to exploring low energy behaviors of correlation functions.

This paper is organized as follows.
In sec.~\ref{sec:2}, we study a correlator which does not have renormalons 
in its OPE components,
after the model is briefly reviewed. We show that after the inverse Laplace transform 
there indeed exists a region where both the OPE and the low energy expansion are simultaneously valid.
We also present how DV appear and how this problem can be circumvented
in the extraction of the low energy expansion coefficients. 
In sec.~\ref{sec:3}, we study a more complictaed correlator in the sense that
it has renormalons in the Wilson coefficients in its OPE, like in the QCD case.
Nevertheless, we show that the calculation and the outcome are almost kept parallel to the first easier example. 
The last section is devoted to the conclusions and discussion.
In appendix \ref{app:A}, we define the Borel transform and the Borel integral,
and also give a brief review which would help the readers who are not familiar with renormalon.
We also present a way to regularize the Borel integral for calculating quantities appearing after the inverse Laplace transform.
In appendix \ref{app:B}, the formulae used for calculating the inverse Laplace transform are given.
Appendix \ref{app:C} explains a supplementary analysis regarding the second example.

\section{$D_{\alpha}(p^2)$: correlation function without renormalons}
\label{sec:2}
 
The action of the $O(N)$ nonlinear $\sigma$ model is given by \cite{Bardeen:1976zh}
\be
S=\frac{1}{2 g_0} \int d^2 x \lt\{ \del_{\mu} \sigma^a(x) \del_{\mu} \sigma^a(x)+\alpha(x)(\sigma^a(x) \sigma^a(x)-N) \rt\}
\ee
where $g_0$ is the bare coupling constant. $\sigma^a(x)$ with $a=1,...,N$ are the dynamical fields 
and $\alpha(x)$ is the Lagrange multiplier to impose $\sum_{a=1}^N \sigma^a \sigma^a=N$.
In the large-$N$ limit, $\alpha$ acquires a non-zero vacuum expectation value (VEV) as
\be
\langle \alpha \rangle=m^2 \equiv \mu^2 \exp[-1/\hat{g}(\mu^2)] ,
\ee
where $\hat{g}(\mu^2)=1/\log{(\mu^2/m^2)}$ is the renormalized coupling in the ${\rm \overline{MS}}$ scheme, related with the original coupling as
\be
\hat{g}_0=\frac{g_0}{4 \pi} .
\ee
$\mu$ denotes the renormalization scale.
The beta function is given by
\be
\mu^2 \frac{d}{d\mu^2} \hat{g}(\mu^2)=-\hat{g}^2(\mu^2),
\ee
indicating that the theory is asymptotically free, and $m$ is the dynamical scale of the model. 
The non-zero VEV generates a mass gap.

In this section, we study the propagator of $\delta \alpha \equiv \alpha-\langle \alpha \rangle$, whose exact result is given by \cite{Novikov:1984ac}
\be
D_{\alpha}(p^2) \equiv N \int d^2 x e^{-i p \cdot (x-y)} \langle \delta \alpha(x) \delta \alpha(y) \rangle
=\frac{4 \pi \sqrt{p^2(p^2+4m^2)}}{\log{\lt[\frac{\sqrt{p^2+4m^2}+\sqrt{p^2}}{\sqrt{p^2+4m^2}-\sqrt{p^2}} \rt]}}  . \label{Dalphaexact}
\ee

\subsection{Low energy and high energy expansions}

The high energy expansion of $D_{\alpha}(p^2)$ is given by \cite{David:1982qv}
\be
D_{\alpha , {\rm OPE}}(p^2)=p^2 \sum_{n=0}^{\infty} C_n(p^2 ) \lt( \frac{m^2}{p^2} \rt)^n , \label{Dhighexp}
\ee
with
\begin{align}
C_0(p^2)&=4 \pi \hat{g}(p^2) , \non
C_1(p^2)&=4 \pi (2 \hat{g}(p^2)-2 \hat{g}^2(p^2)) , \non
C_2(p^2)&=4 \pi (-2 \hat{g}(p^2)-\hat{g}^2(p^2)+4 \hat{g}^3(p^2)) , \non
\vdots
\end{align}
by expanding the exact answer~\eqref{Dalphaexact} in $m^2/p^2$.
The perturbative series of the Wilson coefficients $C_n(p^2)$ terminate at finite orders in $\hat{g}(p)$,
and there are no renormalon divergences.
The power term in $m^2/p^2$ is identified with $(m^2/p^2)^n=\langle \alpha^n \rangle/p^{2n}$, and hence
eq.~\eqref{Dhighexp} is regarded as the OPE.
The Borel transform of $D_{\alpha}(p^2)$ is given by \cite{Hayashi:2023fgl}
\be
B_{D_{\alpha, {\rm OPE}}}(u)=p^2 \sum_{n=0}^{\infty} \lt( \frac{m^2}{p^2} \rt)^n \lt(\frac{\mu^2}{p^2} \rt)^u X_n(u) \label{BDalpha}
\ee
with 
\be
X_n(u)=4 \pi \frac{(4u^2-2u-2n) \Gamma(-n-2u)}{\Gamma(n+1) \Gamma(2-2u-2n)} ,
\ee
where $(\mu^2/p^2)^u X_n(u)$ represents the Borel transform of $C_n(p^2)$.
The definition of the Borel transform is given in appendix~\ref{app:A}.
The OPE~\eqref{Dhighexp} converges for $|p^2|>4m^2$.
We note that high energy expansions--specially, the Wilson coefficients--can be obtained 
not only in solvable models but also in general asymptotically free theories order by order.

\begin{figure}[tb]
\begin{center}
\includegraphics[width=10cm]{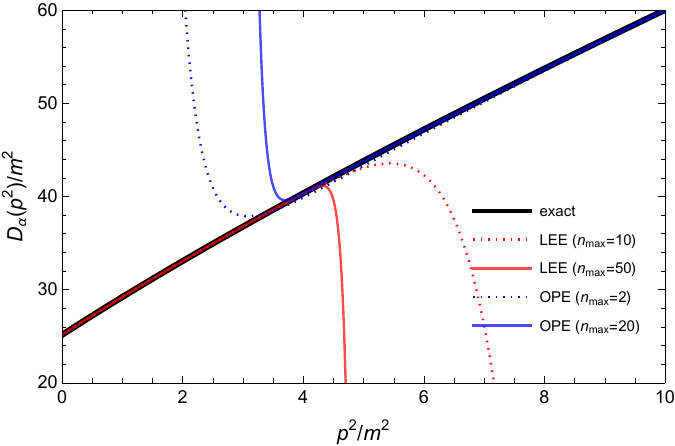}
\end{center}
\caption{Comparison between the exact result~\eqref{Dalphaexact}, 
the low energy expansion (abbreviated as LEE in the figure)~\eqref{lowenergyDalpha},
and the OPE \eqref{Dhighexp} as functions of $p^2/m^2$. $n_{\rm max}$ represents the upper end of the sum of
eqs.~\eqref{Dhighexp} or  \eqref{lowenergyDalpha}.}
\label{figDalpha}
\end{figure}

The low energy expansion is obtained as
\be
D_{\alpha}(p^2)=m^2 \sum_{n=0}^{\infty} c_n \lt(\frac{p^2}{m^2} \rt)^n  \label{lowenergyDalpha}
\ee 
with
\be
c_0=8 \pi, \quad{} c_1=4 \pi/3, \quad{} c_2=-2\pi/45, \cdots,
\ee
by expanding the exact answer~\eqref{Dalphaexact} in $p^2/m^2$.
Because $D_{\alpha}(p^2)$ has the only singularity at $p^2=-4m^2$ ($m_{\rm gap}=2m$),
this expansion converges for $|p^2|<4 m^2$.
Hence there is no region where both the OPE and the low energy expansion are simultaneously valid,
as can be seen from fig.~\ref{figDalpha}.

\subsection{Inverse Laplace transform}

\begin{figure}
\begin{minipage}{0.5\hsize}
\begin{center}
\includegraphics[width=6cm]{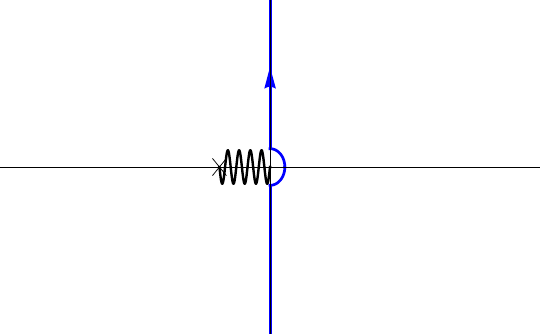}
\end{center}
\end{minipage}
\begin{minipage}{0.5\hsize}
\begin{center}
\includegraphics[width=6cm]{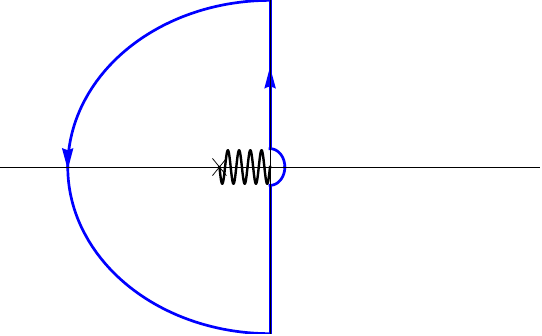}
\end{center}
\end{minipage}
\begin{minipage}{0.5\hsize}
\begin{center}
\includegraphics[width=6cm]{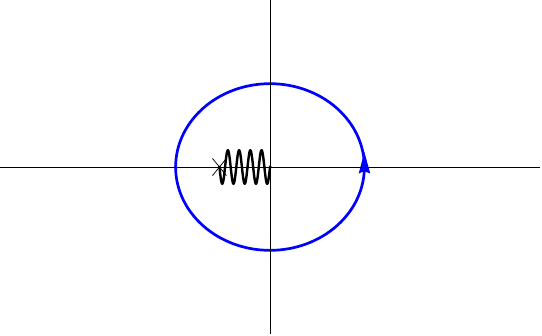}
\end{center}
\end{minipage}
\begin{minipage}{0.5\hsize}
\begin{center}
\includegraphics[width=6cm]{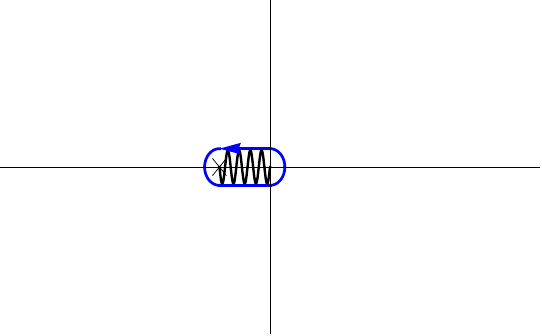}
\end{center}
\end{minipage}
\caption{Integration contours in the complex $r$-plane, where $r=1/p^2$.
The singular points correspond to $p^2=-m_{\rm gap}^2$ ($1/r=-1/m_{\rm gap}^2$), and the cut continues to $p^2=-\infty$ ($r=0$).}
\label{contours}
\end{figure}

We consider the inverse Laplace transform
\be
\tilde{D}_{\alpha}(\tau) \equiv \frac{1}{2\pi i} \int_{-i \infty}^{i \infty} \frac{dr}{r} D_{\alpha}(1/r) e^{\tau r} \label{inverseLaplace}
\ee
where $r \equiv1/p^2$. The mass dimension of $\tau$ is two. 
The integration contour is depicted in the top-left panel of fig.~\ref{contours}.

To obtain the high energy expansion of $\tilde{D}_{\alpha}(\tau)$,
\be
\tilde{D}_{\alpha, {\rm OPE}}(\tau)=\tau \sum_{n=0}^{\infty} \tilde{C}_n(\tau) \lt(\frac{m^2}{\tau} \rt)^n , \label{tildeDalphaOPE}
\ee
it is convenient to calculate its Borel transform, which can be calculated through the Borel transform of $D_{\alpha,{\rm OPE}}(p^2)$:
\begin{align}
B_{\tilde{D}_{\alpha, {\rm OPE}}}(u)
&=\frac{1}{2\pi i}  \int_{-i \infty}^{i \infty} \frac{dr}{r} \lt(B_{D_{\alpha, {\rm OPE}}}(u)|_{p^2 \to 1/r}  \rt)e^{\tau r} \non
&=-\frac{\tau}{\pi} \sum_{n=0}^{\infty} \lt(\frac{m^2}{\tau} \rt)^n  \lt(\frac{\mu^2}{\tau} \rt)^u X_n(u) \sin{[\pi (n+u)]}\Gamma(n+u-1) , \label{tildeDalphaBorel}
\end{align}
where we used a formula in appendix~\ref{app:B} to calculate the inverse Laplace transform.
From this result, the Wilson coefficients $\tilde{C}_n(\tau)$ with small $n$ read
\begin{align}
\tilde{C}_0(\tau)&=4\pi \hat{g}(\tau)-4 \pi (\gamma_E-1)\hat{g}^2(\tau)+\mathcal{O}(\hat{g}^3) \non
\tilde{C}_1(\tau)&=8 \pi \hat{g}(\tau)-8 \pi (\gamma_E+1)\hat{g}^2(\tau)+\mathcal{O}(\hat{g}^3) \non
\vdots
\end{align}
Although the perturbative series of the Wilson coefficients do not terminate at finite orders,
there are no renormalon divergences as one can see from the fact that (each $n$th term of)
the Borel transform~\eqref{tildeDalphaBorel} does not
have singularities in the Borel $u$-plane.
Note that the sine function cancels the poles of the gamma function.

On the other hand, the low energy expansion is given by
\be
\tilde{D}_{\alpha}(\tau)=m^2 \sum_{n=0}^{\infty} \frac{c_n}{n!} \lt(\frac{\tau}{m^2} \rt)^n \label{tildeDalphalow}
\ee 
where $c_n$ is defined in eq.~\eqref{lowenergyDalpha}.
The above series expansion is derived as follows.
Since the low energy behavior of $D_{\alpha}$ is given by $D_{\alpha}(1/r) \sim r^0$ as $r \to \infty$,
the contribution from the large arc in the integral contour shown in the top-right panel of fig.~\ref{contours} vanishes.
Then the integral along the original contour~\eqref{inverseLaplace} is equivalent to 
that along the contour shown in the bottom-left panel, on which the low energy expansion~\eqref{lowenergyDalpha} is valid.
The residue theorem then leads to eq.~\eqref{tildeDalphalow}.
The low energy expansion~\eqref{tildeDalphalow} can 
be considered an exact expression of $\tilde{D}_{\alpha}(\tau)$.
This is because the convergence radius of eq.~\eqref{tildeDalphalow} is infinity 
and thus $\tilde{D}_{\alpha}(\tau)$ can be defined as an analytic function through eq.~\eqref{tildeDalphalow}.
The fact that the series has an infinite convergence radius can be proven based on the Cauchy-Hadamard theorem and the fact that 
eq.~\eqref{lowenergyDalpha} has a non-zero convergence radius.

\begin{figure}
\begin{center}
\includegraphics[width=9cm]{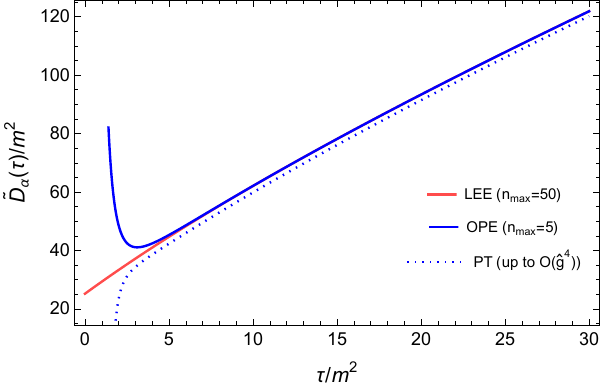}
\end{center}
\caption{Comparison between the low energy expansion (LEE)~\eqref{tildeDalphalow}
and the OPE~\eqref{tildeDalphaOPE} for $\tilde{D}_{\alpha}(\tau)/m^2$.
PT stands for perturbation theory and means the perturbative series of $(\tau/m^2)\tilde{C}_0(\tau)$ of eq.~\eqref{tildeDalphaOPE}.
The perturbative series truncated at $\hat{g}(\tau)^4$ is shown.
$n_{\rm max}$ represents the upper limit of the sum of the corresponding expressions.}
\label{figtildeDalpha}
\end{figure}

The high energy expansion $\tilde{D}_{\alpha, {\rm OPE}}(\tau)$ \eqref{tildeDalphaOPE} is compared with the low energy expansion~\eqref{tildeDalphalow} in fig.~\ref{figtildeDalpha}.
The leading term of the OPE, i.e., $\tau C_0(\tau)$, corresponds to 
the usual perturbative evaluation of $\tilde{D}_{\alpha}(\tau)$
and  the series truncated at $\mathcal{O}(\hat{g}^4)$ is shown in the figure.
The OPE result (the blue line) truncated at $n_{\rm max}=5$
is given, where we use the all-order perturbative results for the Wilson coefficients $\tilde{C}_{0-5}(\tau)$, obtained by the Borel integral.
Unless otherwise stated, we use the all-order result for Wilson coefficients in the following figures. 
One can find a region where both the OPE and the low energy expansion are valid, 
thanks to the enlargement of the convergence radius of the low energy expansion.

We make a remark on how the Borel integral is calculated in giving the all-order perturbative result of $\tilde{C}_n(\tau)$.
Although there are no renormalons in $\tilde{C}_n(\tau)$,
we need to modify the integration contour of the Borel integral. 
This is because of a non-convergent behavior at $u \to \infty$
due to the factor $\Gamma(n+u-1) \sim e^{u \log u-u}$, arising from the inverse Laplace transform.
We regularize the Borel integral by rotating its path to the positive imaginary axis or the negative imaginary axis
after rewriting the sine factor in eq.~\eqref{tildeDalphaBorel} as 
$\sin{[\pi (n+u)]}=(e^{i \pi (n+u)}-e^{-i \pi (n+u)})/(2i)$.
In appendix~\ref{app:A}, we explain the detailed calculation method  
and how the regularization is justified in the context of a more proper calculation of $\tilde{D}_{\alpha}(\tau)$.
We use eq.~\eqref{Boreintregularize} in practice.

\subsection{Duality violations}
\label{sec:2.3}

\begin{figure}[tb]
\hspace{2.5cm}
\includegraphics{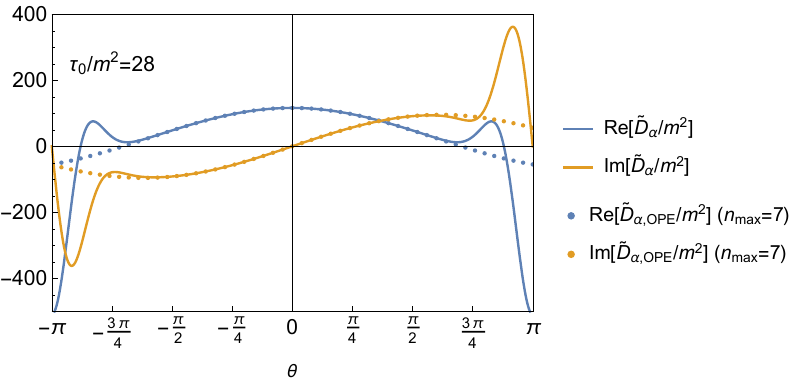}
\caption{Comparison between the OPE~\eqref{tildeDalphaOPE} truncated at $n_{\rm max}=7$ 
and the exact result on the complex circle $\tau=\tau_0 e^{i \theta}$ with $\tau_0/m^2=28$.
The exact results (solid lines) are obtained by eq.~\eqref{tildeDalphalow}, where the series is truncated at a sufficiently large order.
The OPE results are represented by the blue points (real part) and the orange points (imaginary part).
}
\label{figtildeDalphaOPEComplexcircle}
\end{figure}

Although the OPE for $\tilde{D}_{\alpha}(\tau)$ is shown to well approximate
the low energy expansion~\eqref{tildeDalphalow} for positive real $\tau \gg m^2$ in fig.~\ref{figtildeDalpha}, 
the OPE actually deviates significantly from the exact result on a complex circle $\tau=\tau_0 e^{i \theta}$ ($\tau_0 >0$, $-\pi < \theta < \pi $)
when ${\rm Re} \, [\tau] <0$ even for $\tau_0 \gg m^2$.
As an example, the OPE truncated at $n_{\rm max}=7$ is compared with the exact result
on a complex circle with $\tau_0/m^2=28$ in fig.~\ref{figtildeDalphaOPEComplexcircle}.
We call such deviation duality violations (DV). 

The origin of the DV can be understood as follows.
The exact result of $\tilde{D}_{\alpha}(\tau)$ can also be obtained by the integral along the path of the bottom-right panel of fig.~\ref{contours},
where the exact result has non-zero imaginary part from $r=0$ to $r=-1/(4m^2)$.
However, the OPE truncated at a finite order has non-zero imaginary part from $r=0$ to $r=-\infty$.
One can confirm that the OPE result \eqref{tildeDalphaOPE} with eq.~\eqref{tildeDalphaBorel}
is equivalent to the integral of the imaginary part along $r=-\infty$ to $r=0$:\fn{
Strictly speaking, since the integral around $r=0$ diverges, we need to consider the contribution from the small arc around $r=0$.
After this is considered the result gets finite and the result of eq.~\eqref{tildeDalphaBorel} can be obtained.\label{fn2}}
\be
\tilde{D}_{\alpha, {\rm OPE}}(\tau)=\frac{1}{\pi} \int_{0+i0}^{-\infty+i0} \frac{dr}{r} {\rm Im} \, [D_{\alpha, {\rm OPE}}(1/r) ] e^{\tau r} . \label{tildeDalphaOPEanother}
\ee
Therefore, the relation of the exact result to the OPE is given by
\be
\tilde{D}_{\alpha}(\tau)=\tilde{D}_{\alpha, {\rm OPE}}(\tau)+\tilde{D}_{\alpha, {\rm DV}}(\tau) , \label{OPEplusDV}
\ee
with
\be
\tilde{D}_{\alpha, {\rm DV}}(\tau)=-\frac{1}{\pi} \int_{-1/(4m^2)+i0}^{-\infty+i0} \frac{dr}{r} {\rm Im} \, [ D_{\alpha,  {\rm OPE}}(1/r) ]e^{\tau r} . \label{tildeDalphaDV}
\ee
One can see that the DV originates from the use of the OPE in the energy region ($-4m^2 < p^2 <0$) where its use is not valid.

We investigate an approximate function of $\tilde{D}_{\alpha, {\rm DV}}(\tau)$.
To this end, we first consider the Borel transform of $D_{\alpha, {\rm DV}}$:
\begin{align}
B_{\tilde{D}_{\alpha,{\rm DV}}}(u) 
&\equiv 
-\frac{1}{\pi} \int_{-1/(4m^2)+i0}^{-\infty+i0} \frac{dr}{r}  {\rm Im} \lt[ B_{D_{\alpha,  {\rm OPE}}}(u)|_{p^2 \to 1/r} \rt]e^{\tau r} \non
&=\frac{\tau}{\pi} \sum_{n=0}^{\infty} \lt(\frac{m^2}{\tau} \rt)^n  \lt(\frac{\mu^2}{\tau} \rt)^u X_n(u) \sin[\pi (n+u)] \Gamma(n+u-1, \tau/(4 m^2)) , \label{tildeDalphaDVBorel}
\end{align}
where $\Gamma(a,z)=\int_z^{\infty} dx \, x^{a-1} e^{-x}$ is the incomplete gamma function.
We note that, although one can formally obtain the high energy expansion of $\tilde{D}_{\alpha}$ from this result,
it does not give a systematic expansion in $\hat{g}$ or $m^2/\tau$.
This can be seen by noting the asymptotic expansion of the incomplete gamma function for $z \gg 1$,
\be
\Gamma(a,z)=e^{-z} z^{a-1} \lt[1+\frac{a-1}{z}+\frac{(a-1)(a-2)}{z^2}+\cdots \rt] . \label{incompGammaapprox}
\ee
The $k$th order perturbative coefficient is obtained by $(d^k/d u^k) B_{\tilde{D}_{\alpha,{\rm DV}}}(u)|_{u \to 0}$,
which then has $\log^k{[\tau/(4m^2)]} \sim (1/\hat{g}(\tau))^k$. 
One can also see that a larger $n$ contribution in eq.~\eqref{tildeDalphaDVBorel} is not always small in terms of $m^2/\tau \ll 1$ 
compared with a smaller $n$ contribution.
These features are not very surprising because $\tilde{D}_{\alpha, {\rm DV}}$ is given through the OPE at low energy region, where the OPE is not valid, as already mentioned. 
$\tilde{D}_{\alpha,{\rm DV}}$ can be approximated instead using the Borel integral:
\begin{align}
\tilde{D}_{\alpha, {\rm DV}}(\tau)
&=\int_0^{\infty} du\, B_{\tilde{D}_{\alpha,{\rm DV}}}(u) e^{-u/\hat{g}(\mu^2)} \non
&\approx   \frac{16 m^4}{\pi \tau} e^{-\tau/(4 m^2)} \sum_{n=0}^{\infty} 4^{-n} \int_0^{\infty} du\, X_n(u) \sin[\pi (n+u)]  4^{-u} \non
&\quad{} \times  \lt[1+(n+u-2) \lt(\frac{4m^2}{\tau} \rt)+(n+u-2)(n+u-3)  \lt(\frac{4m^2}{\tau} \rt)^2+\cdots  \rt] , \non
&=K \frac{e^{-\tau/(4m^2)}}{\tau} \lt[1+k_1 \lt(\frac{4m^2}{\tau} \rt)+k_2 \lt(\frac{4m^2}{\tau} \rt)^2+\cdots   \rt] \label{tildeDalphaDVapprox}
\end{align}
where we used eq.~\eqref{incompGammaapprox} to approximate $B_{\tilde{D}_{\alpha, {\rm DV}}}(u)$.
$K, k_1, k_2, \dots$ denote $\tau$-independent constants.
We  note that, since the DV are correction to be added to {\it{a truncated OPE}}, 
one should practically truncate the sum over $n$ in eq.~\eqref{tildeDalphaDVapprox} in accordance with the OPE truncation order.
While the approximate function is always given in the form of eq.~\eqref{tildeDalphaDVapprox}
regardless of the truncation order, the constants $K, k_1, k_2,\dots$ depend on the truncation order. 
We  also note that the above function form does not depend much on the details of the observable under consideration.
The above function form is essentially determined by the incomplete gamma function,
which originates from the integration range of eq.~\eqref{tildeDalphaDVBorel}.

\begin{figure}[tb]
\hspace{2.5cm}
\includegraphics{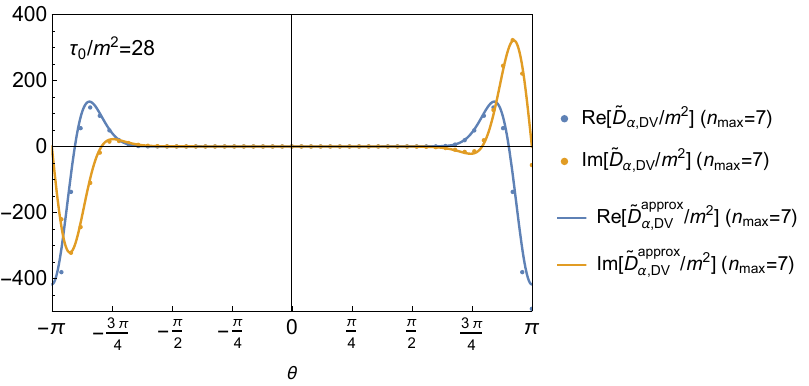}
\caption{Comparison between $\tilde{D}_{\alpha, {\rm DV}}(\tau)$ and its approximation formula~\eqref{tildeDalphaDVapprox} 
on the complex circle $\tau=\tau_0 e^{i \theta}$ with $\tau_0/m^2=28$.
See the main text for the details. We take the first three terms of the square brackets in eq.~\eqref{tildeDalphaDVapprox}
for $\tilde{D}^{\rm approx}_{\alpha, {\rm DV}}$.}
\label{figtildeDalphaDV}
\end{figure}

In fig.~\ref{figtildeDalphaDV}, 
we compare $\tilde{D}_{\alpha, {\rm DV}}(\tau)$ of eq.~\eqref{tildeDalphaDV} truncating $D_{\alpha, {\rm OPE}}(1/r)$ at $n_{\rm max}=7$ 
(labeled as ``${\rm Re} [\tilde{D}_{\alpha,{\rm DV}}/m^2]~(n_{\rm max}=7)$'' and ``${\rm Im} [\tilde{D}_{\alpha,{\rm DV}}/m^2]~(n_{\rm max}=7)$'' in the figure)
with the approximate function of eq.~\eqref{tildeDalphaDVapprox} 
(labeled as ``${\rm Re} [\tilde{D}^{\rm approx}_{\alpha,{\rm DV}}/m^2]~(n_{\rm max}=7)$'' and ``${\rm Im} [\tilde{D}^{\rm approx}_{\alpha,{\rm DV}}/m^2]~(n_{\rm max}=7)$'').
In practice, we give $\tilde{D}_{\alpha, {\rm DV}}(\tau)$ based on the Borel integral, i.e., 
the right-hand side of the first equality of eq.~\eqref{tildeDalphaDVapprox} with truncating the sum over $n$ in $B_{\tilde{D}_{\alpha,{\rm DV}}}(u)$.
Since the Borel integral again shows the divergent behavior at $u \to \infty$ 
(as the asymptotic behavior of $\Gamma(u,z)$ at large $u$ is the same as $\Gamma(u)$),
we use eq.~\eqref{Boreintregularize} to regularize the Borel integrals.
For $\tilde{D}^{\rm approx}_{\alpha,{\rm DV}}$ in the figure, we take the first three terms inside the square brackets in eq.~\eqref{tildeDalphaDVapprox}.

We make comments related to the DV.
The fact that the OPE $\tilde{D}_{\alpha, {\rm OPE}}(\tau)$ needs the exponentially suppressed corrections $\sim e^{-\tau/(4m^2)}$
in terms of its expansion parameter $m^2/\tau \ll 1$ suggests a possibility that the OPE is a divergent asymptotic expansion.
We numerically observed that the $\tau$-space OPE (for a real positive $\tau$) indeed exhibits a divergent behavior at large orders
in the $m^2/\tau$-expansion
and the OPE term reaches a minimum at $n \sim \tau/(4 m^2)$. 
This relation is in agreement with a theoretical consideration on the interplay between an asymptotic series and exponentially suppressed corrections; 
if an asymptotic series $\sum_n a_n x^{n+1}$ has the asymptotic behavior of $a_n \sim n!/u_0^n$ with a positive parameter $u_0$,
the minimum term of the asymptotic series is obtained at $n \sim u_0/ x$ and a nonperturbative correction of $e^{-u_0/x}$  appears.
We also numerically confirmed that the sum of the OPE and the DV (eq.~\eqref{OPEplusDV}) indeed converges to the exact value,
as expected from the convergence of $D_{\alpha, {\rm OPE}}(1/r)$ for $-1/(4m^2) < r \leq 0 $.

In the subsequent subsection, we exclusively rely on $\tilde{D}_{\alpha, {\rm OPE}}(\tau)$ 
to extract the low energy limit  $\lim_{p^2 \to 0}D_{\alpha}(p^2)$
without directly employing the explicit result of $\tilde{D}_{\alpha, {\rm DV}}(\tau)$.
This decision is because $\tilde{D}_{\alpha, {\rm OPE}}(\tau)$ can be calculated systematically in general asymptotically free theories.
However, we will implicitly use the information that the DV are given by $\sim \frac{m^4}{\tau} e^{-\tau/(4 m^2)}$.
As noted, it is expected that the function form does not depend much on observables under consideration.
Also the exponent can be deduced from the mass gap ($m_{\rm gap}=2m$ in this case)
or, notably, from the divergent behavior of 
the asymptotic expansion of $\tilde{D}_{\alpha, {\rm OPE}}(\tau)$, as mentioned in the previous paragraph.
The latter can be calculated within a systematic short distance expansion.

\subsection{Low energy limit from the OPE}

Now we extract the low energy limit $D_{\alpha}(p^2=0)/m^2=c_0=8\pi$ using the OPE 
in the $\tau$-space, $\tilde{D}_{\alpha ,{\rm OPE}}(\tau)$.
However, employing $\tilde{D}_{\alpha, {\rm OPE}}(\tau)$ as an approximation for $\tilde{D}_{\alpha}(\tau)$ 
in the formula
\be
c_0=\frac{1}{2 \pi i} \oint_{\tau=\tau_0 e^{i \theta}} \frac{d \tau}{\tau} \tilde{D}_{\alpha}(\tau)/m^2  \quad{} {\text{with $\tau_0>0$}} ,
\ee
where the integration contour surrounds the origin by the circle of radius $\tau_0$,
is not valid because of the significant DV; 
the approximation $\tilde{D}_{\alpha}(\tau) \approx \tilde{D}_{\alpha, {\rm OPE}}(\tau)$ becomes unreliable for ${\rm Re} [\tau]<0$. 
We then consider the following integral where the negative-${\rm Re} [\tau]$ region is suppressed 
\be
c_0=\frac{1}{2 \pi i} \oint_{\tau=\tau_0 e^{i \theta}} \frac{d \tau}{\tau} \lt(\frac{\tau}{\tau_0}+1 \rt)^k \tilde{D}_{\alpha}(\tau)/m^2 ,
\ee
where $k$ is an arbitrary positive integer.
Based on this formula, we extract $c_0$ using $\tilde{D}_{\alpha, {\rm OPE}}(\tau)$:
\be
c_0 \approx \frac{1}{2 \pi i} \oint_{\tau=\tau_0 e^{i \theta}} \frac{d \tau}{\tau} \lt(\frac{\tau}{\tau_0}+1 \rt)^k \tilde{D}_{\alpha, {\rm OPE}}(\tau)/m^2 . \label{c0withk}
\ee
In this formula the DV of form eq.~\eqref{tildeDalphaDVapprox} can be suppressed by choosing $k \sim \tau_0/(4 m^2)$
as long as $m^2/\tau_0 \ll 1$
because
\begin{align}
&\frac{1}{2 \pi i} \oint_{\tau=\tau_0 e^{i \theta}} \frac{d \tau}{\tau} \lt(\frac{\tau}{\tau_0}+1 \rt)^k \frac{m^2}{\tau} e^{-\tau/(4m^2)} 
=\frac{m^2}{\tau_0} \lt(k-\frac{\tau_0}{4m^2} \rt) , \non
&\frac{1}{2 \pi i} \oint_{\tau=\tau_0 e^{i \theta}} \frac{d \tau}{\tau} \lt(\frac{\tau}{\tau_0}+1 \rt)^k \lt( \frac{m^2}{\tau} \rt)^2 e^{-\tau/(4m^2)}
=-\frac{m^2}{8 \tau_0}+\mathcal{O}\lt(k-\frac{\tau_0}{4m^2} \rt)   , \non
&\frac{1}{2 \pi i} \oint_{\tau=\tau_0 e^{i \theta}} \frac{d \tau}{\tau} \lt(\frac{\tau}{\tau_0}+1 \rt)^k \lt( \frac{m^2}{\tau} \rt)^3 e^{-\tau/(4m^2)}
=\frac{m^4}{12 \tau_0^2}+\mathcal{O}\lt(k-\frac{\tau_0}{4m^2} \rt) , \non
&\qquad{}\vdots \label{suppressDV}
\end{align}

\begin{figure}[tb]
\begin{minipage}{0.5\hsize}
\begin{center}
\includegraphics[width=7cm]{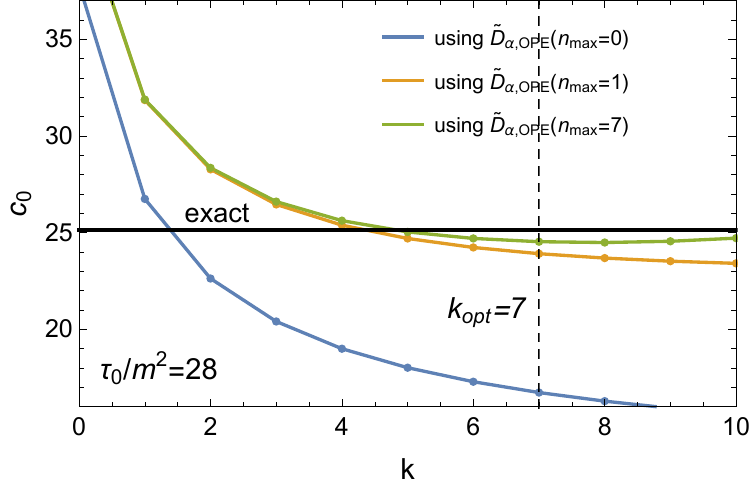}
\end{center}
\end{minipage}
\begin{minipage}{0.5\hsize}
\begin{center}
\includegraphics[width=7cm]{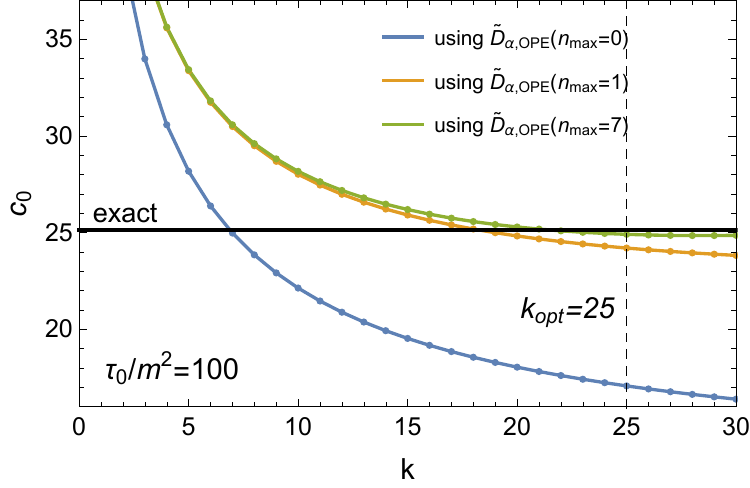}
\end{center}
\end{minipage}
\caption{Extraction of $c_0$ based on eq.~\eqref{c0withk} using $\tilde{D}_{\alpha, {\rm OPE}}(\tau)$. The radius of the integral contour is
$\tau_0/m^2=28$ (left) and $\tau/m^2=100$ (right), respectively. 
The horizontal axis shows $k$ in eq.~\eqref{c0withk} and
the point corresponding to the optimal choice $k_{\rm opt}=\tau_0/(4m^2)$ is shown in each case.
The truncation orders of the OPE (denoted by $n_{\rm max}$) are indicated in the figures. 
The exact answer $c_0=8 \pi$ is shown by the black lines.}
\label{figc0Dalpha}
\end{figure}

The result of $c_0$ extracted based on eq.~\eqref{c0withk} is shown in fig.~\ref{figc0Dalpha}.
We take the radius of the integration contour as $\tau_0/m^2=28$ (left) and $\tau_0/m^2=100$ (right),
where the optimal $k$ is given by $k_{\rm opt}=7$ and $k_{\rm opt}=25$, respectively.
In the figure, the results with different OPE truncation orders are shown.
One can see that the truncated OPE with $n_{\rm max}=1$ already gives 
an estimate with precision about 5 \% with the optimal choices of $k$.
We infer that the reason why the OPE result is already stable at $n_{\rm max}=1$ is because 
the Wilson coefficients $\tilde{C}_n(\tau)$ with $n \geq 2$ are suppressed as $\mathcal{O}(\hat{g}^2)$, 
while $\tilde{C}_{0, 1}(\tau) \sim \mathcal{O}(\hat{g})$. 

\begin{figure}[tb]
\begin{center}
\includegraphics[width=8cm]{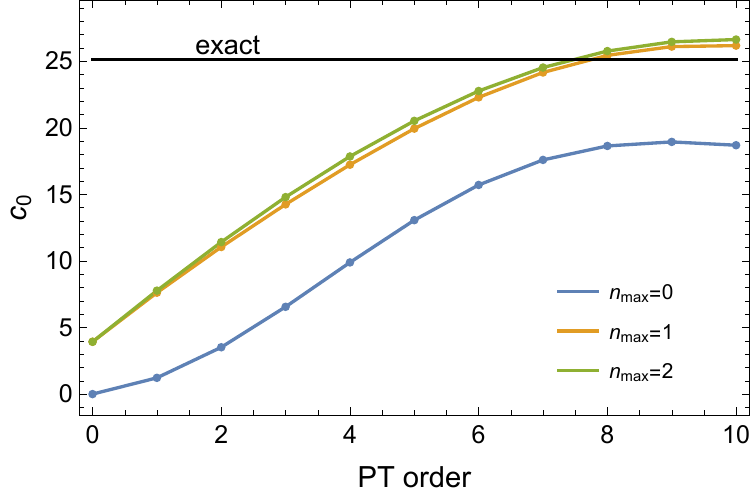}
\end{center}
\caption{Extraction of $c_0$ based on eq.~\eqref{c0withk} with $\tau_0/m^2=600$, using 
fixed order perturbation theory for the Wilson coefficients.
The blue line is the estimate where the OPE is truncated at $n_{\rm max}=0$
and the Wilson coefficient $\tilde{C}_0(\tau)$ is calculated by fixed order perturbation theory;
the order of the perturbation theory is indicated by ``PT order''. 
(``PT order''$=i$ corresponds to truncating the series at $\mathcal{O}(\hat{g}^{i+1})$.)
The orange and green lines are obtained for the OPE truncation order at $n_{\rm max}=1,2$, respectively.
For all the involved Wilson coefficients $\tilde{C}_n(\tau)$, the perturbative series are truncated 
at the same order indicated by ``PT order''.
$k$ in eq.~\eqref{c0withk} is fixed to the optimal value $k=\tau_0/(4m^2)=150$.
The exact answer $c_0=8 \pi$ is shown by the black lines.}
\label{figc0Dalpha600PT}
\end{figure}

Keeping more practical applications in mind, we also show a result 
using fixed order perturbation theory for the Wilson coefficients in $\tilde{D}_{\alpha, {\rm OPE}}$, 
rather than using the Borel integral.
In fig.~\ref{figc0Dalpha600PT}, we show the result with $\tau_0/m^2=600$,
fixing $k$ to its optimal value $k=600/4=150$.
The horizontal axis represents the truncation order of the perturbative series for the Wilson coefficients $\tilde{C}_n(\tau)$,
while the truncation order of the OPE is fixed for each colored line.
We take the renormalization scale $\mu$ as $\mu^2=\tau_0=600 m^2$
for the perturbative series.
It is seen that the estimates indeed approach the exact value as the order of perturbation theory and that of OPE are raised.
The OPE higher orders beyond $n_{\rm max}=1$ give minor corrections, as already expected from fig.~\ref{figc0Dalpha}.

\section{$\rho(q^2)$: correlation function with renormalons}
\label{sec:3}

In this section, as a second example, we consider the self-energy correction to the $\sigma$-propagator at $1/N$ \cite{Beneke:1998eq}.
At the $1/N^0$ order, the $\sigma$ propagator is given by the standard form $\sim 1/(q^2+m^2)$. 
At the $1/N$ order, it has a self-energy correction, given by
\be
\frac{1}{N} \Pi_{\sigma}(q^2)=\frac{1}{N}\int \frac{d^d k}{(2 \pi)^d} \frac{1}{(q+k)^2+m^2} D_{\alpha}(k^2) \label{Pi}
\ee
with $d=2-2 \epsilon$.\fn{
Although we employ $D_{\alpha}(k^2)$ that is obtained for $d=2$, the results presented below would not be changed if 
$D_{\alpha}(k^2)$ for $d=2-2\epsilon$ is used because a finite quantity is studied.}
Since this is UV divergent, we instead study a UV finite quantity,
\be
\rho(q^2) \equiv q^2 \frac{d^2}{d (q^2)^2} \Pi_{\sigma}(q^2) . \label{rhodef}
\ee
which is defined to be dimensionless.
A distinct difference from the first example is that the Wilson coefficients in the OPE of $\rho(q^2)$
have renormalon divergences like in the QCD case.

We note that the symbols of $c_n$ and $\tilde{C}_n(\tau)$ are renewed in this section, 
thus they are defined as different quantities from those in sec.~\ref{sec:2}.

\subsection{Low energy and high energy expansions}

The OPE of $\Pi_{\sigma}(q^2)$ was studied in detail in ref.~\cite{Beneke:1998eq},
and that of $\rho(q^2)$ has been also calculated in ref.~\cite{Hayashi:2023fgl}.
The OPE is given by
\be
\rho_{\rm OPE}(q^2)=\sum_{n=0}^{\infty} \lt[ C^{1}_n(q^2) \frac{\langle \alpha^n \rangle}{q^{2n}}+
\sum_{j} C^{2}_{n,j}(q^2) \frac{\langle \mathcal{O}^{\delta \alpha}_{n,j} \rangle}{q^{2n}} \rt] \label{rhoOPE}
\ee
where $\langle \alpha^n \rangle=m^{2n}$ and $\langle \mathcal{O}^{\delta \alpha}_{n,j} \rangle$ denotes the matrix element of 
a local operator of mass dimension $2n$ which  involves $\delta \alpha$, 
e.g., the lowest operator is given by $\langle \mathcal{O}^{\delta \alpha}_{n=2} \rangle=\langle \delta \alpha(x) \delta \alpha(x) \rangle$;
for $n=0, 1$ the $\langle \mathcal{O}^{\delta \alpha}_{n,j} \rangle$ term vanishes.
The Wilson coefficients $C_n^1(q^2)$ generally have renormalons and  
the matrix elements $\langle \mathcal{O}^{\delta \alpha}_{n,j} \rangle$ are ambiguous due to their UV divergences.
It was shown in ref.~\cite{Beneke:1998eq} that these ambiguities are completely canceled in the OPE
and that the OPE (the sum over $n$) converges for $|q^2|>9m^2$.
$9 m^2$ corresponds to the mass gap in this channel $m_{\rm gap}=m+2m=3m$,
where $m^2$ is the pole of the $1/N^0$-order $\sigma$ propagator and 
$(2m)^2$ is the singularity of $D_{\alpha}(k^2)$.

The Borel transform of $\sum_{n=0}^{\infty} C^{1}_n(q^2) \lt( \frac{ m^2 }{q^{2}} \rt)^n$ is given by \cite{Beneke:1998eq,Hayashi:2023fgl}\fn{
The Borel transform of $C_n^1(q^2)$ is understood as $\frac{1}{4 \pi} \lt(\frac{\mu^2}{q^2} \rt)^u [c_n(u) \log{(q^2/m^2)}+d_n(u)]$ as before.
}
\be
B_{\rho_{{\rm OPE}}}(u)=\frac{1}{4 \pi} \lt(\frac{\mu^2}{q^2} \rt)^u \sum_{n=0}^{\infty} \lt(\frac{m^2}{q^2} \rt)^n [c_n(u) \log{(q^2/m^2)}+d_n(u)] , \label{rhoBorel}
\ee
where
\be
c_n(u)=\sum_{j=0}^{n} \frac{(-1)^{n-j}}{((n-j)!)^2} X_j(u) [(j+u-1)_{n-j}]^2 (n+u)(n+u-1) , \label{c}
\ee
and 
\be
d_n(u)=d_n^1(u)-\frac{\pi \cos{(\pi u)}}{\sin{(\pi u)}} d_n^2(u) , \label{ddecom}
\ee
with
\begin{align}
d_n^1(u)
&=\sum_{j=0}^{n} \frac{(-1)^{n-j}}{((n-j)!)^2} X_j(u) [(j+u-1)_{n-j}]^2 \non
&\quad{} \times \{1-2(n+u)+2(n+u)(n+u-1) [\psi(1+n-j)-\psi(n+u-1)] \} , \non
d_n^2(u)
&=\sum_{j=0}^{n} \frac{(-1)^{n-j}}{((n-j)!)^2} X_j(u) [(j+u-1)_{n-j}]^2 (n+u)(n+u-1)(=c_n(u)) . \label{d1d2}
\end{align}
$(a)_c \equiv \Gamma(a+c)/\Gamma(a)$ is the Pochhammer symbol and 
$\psi(z)\equiv\Gamma'(z)/\Gamma(z)$ is the digamma function.
See refs.~\cite{Beneke:1998eq,Hayashi:2023fgl} for the detailed calculation.
Renormalons (singularities in the Borel $u$-plane) arise from 
$1/\sin{(\pi u)}$ in eq.~\eqref{ddecom} and from $\psi(n+u-1)$ in $d_n^1(u)$ of eq.~\eqref{d1d2}.
The digamma function $\psi(n+u-1)$ gives only negative singularities (UV renormalons).
$c_n(u)$ does not have renormalons.
For example, one can see that $C_{n=0}^1(q^2)$ has IR renormalons at $u=2,3,4,\dots$ and UV renormalons at $u=-1,-2,-3\dots$
(The $u=0, 1$ singularities are canceled by the factor $(n+u)(n+u-1)=u(u-1)$.)
Therefore, perturbative calculations of the Wilson coefficients suffer from inevitable uncertainties caused by the IR renormalons.
(Note that this is independent of the renormalon cancellation in the OPE, mentioned above. 
The cancellation occurs when the all-order perturbative series are resummed
and the uncertainties of the nonperturbative matrix elements are explicitly addressed.
IR renormalon uncertainties persist in perturbative evaluations.)
In the present study, we do not need the explicit result of $C_{n,j}^2(q^2)$.
This is because it is known that $C_{n,j}^2(q^2)$ is a $q^2$-independent constant \cite{Beneke:1998eq,Hayashi:2023fgl},
and the $\langle \mathcal{O}_{n,j}^{\delta \alpha} \rangle$ terms disappear after the inverse Laplace transform is applied,
as can be seen from eq.~\eqref{inverseLaplaceformula}.

The exact evaluation of $\rho(q^2)$ can be done by
\begin{align}
\rho(q^2)
&=\frac{q^2}{4 \pi^2}  \int_0^{\infty} \frac{d k^2}{2} D_{\alpha}(k^2) \frac{\del^2}{\del (q^2)^2} \lt[ \int_0^{2\pi} d \theta \frac{1}{q^2+2 q k \cos{\theta}+k^2+m^2} \rt] \non
&=\frac{q^2}{4 \pi^2} \int_0^{\infty} \frac{d k^2}{2} D_{\alpha}(k^2) \frac{\del^2}{\del (q^2)^2} \frac{2 \pi}{[(k^2)^2+2k^2(m^2-q^2)+(m^2+q^2)^2]^{1/2}} . \label{rhoexact}
\end{align}
where the $k^2$-integral is numerically evaluated.
The low energy expansion
\be
\rho(q^2)=\sum_{n=1}^{\infty} c_n \lt( \frac{q^2}{m^2} \rt)^n
\ee
converges for $|q^2| < 9 m^2$ because the closest singularity is located at $q^2=(m_{\rm gap})^2=9m^2$.
The coefficients $c_n$ can be obtained by taking derivatives of eq.~\eqref{rhoexact} 
with respect to $q^2$ and then sending $q^2 \to 0$ before the $k^2$-integral.\fn{
The integrand of the $k^2$-integral to obtain $c_n$ for higher $n$ exhibits increased oscillatory
 behavior around $k^2=0$, which complicates accurate numerical integration, despite its analyticity at $k^2=0$.
We then split the integral as $\int_0^{\infty} dk^2=\lt(\int_0^{\delta} + \int_{\delta}^{\infty} \rt) dk^2$ with $0<\delta <1$
and perform the former integral analytically, expanding the integrand around $k^2=0$ to sufficiently high orders.
}
They are given by
\be
c_1=-0.012396..., \quad{} c_2=0.001437..., \quad{} c_3=-0.00016138..., \quad{} \cdots \label{cnrho}
\ee
The leading expansion coefficient $c_0$ is zero because $\rho(q^2=0)=0$ due to the overall factor $q^2$ in eq.~\eqref{rhodef}.

\subsection{Inverse Laplace transform}

We consider the inverse Laplace transform
\be
\tilde{\rho}(\tau) \equiv \frac{1}{2 \pi i} \int_{-i\infty}^{i \infty} \frac{dr}{r} \rho(1/r) e^{\tau r} .
\ee
with $r=1/q^2$. The OPE of the inverse Laplace transform of $\tilde{\rho}(\tau)$ has been calculated in ref.~\cite{Hayashi:2023fgl}.
(However, we note that slightly different inverse Laplace transforms are considered;
$\int_{- i \infty}^{i \infty} \frac{dr}{r^{\kappa}}$ with $\kappa=2$ is consider in ref.~\cite{Hayashi:2023fgl}
but with $\kappa=1$ here.) 
To obtain the OPE,
\be
\tilde{\rho}_{\rm OPE}(\tau)=\sum_{n=0}^{\infty} \tilde{C}_n(\tau) \lt( \frac{m^2}{\tau}\rt)^n , \label{tilderhoOPE}
\ee
we calculate its Borel transform as before:
\begin{align}
B_{\tilde{\rho}_{\rm OPE}}(u)
&=\frac{1}{2 \pi i}  \int_{-i\infty}^{i \infty} \frac{dr}{r} \lt( B_{\rho_{\rm OPE}}(u)|_{q^2 \to 1/r} \rt)e^{\tau r}  \non
&=\frac{1}{4 \pi^2} \lt(\frac{\mu^2}{\tau} \rt)^u \sum_{n=0}^{\infty} \lt(\frac{m^2}{\tau} \rt)^n \non
&\quad{} \times \big\{ \lt[c_n(u) (\log{(\tau/m^2)}-\psi(u+n))+d_n^1(u) \rt] \Gamma(u+n) \sin{[\pi (u+n)]} \non
&\quad{}\quad{} -(c_n(u)+d_n^2(u)) \pi \Gamma(u+n) \cos{[\pi (u+n)]} 
 \big\} . \label{tilderhoBorel}
\end{align}
Here the formulae eqs.~\eqref{inverseLaplaceformula} and \eqref{inverseLaplaceformula2} are used.
Since $1/\sin{(\pi u)}$ in eq.~\eqref{ddecom} is canceled, one sees that 
$\tilde{C}_n(\tau)$ does not have IR renormalons at all.
At the same time, the ambiguous nonperturbative matrix elements $\langle \mathcal{O}_{n,j}^{\delta \alpha} \rangle$
do not appear in $\tilde{\rho}_{\rm OPE}(\tau)$, as mentioned before.
Therefore, $\tilde{C}_n(\tau)$ can be calculated accurately even by usual fixed order perturbation theory.
These features are already clarified in ref.~\cite{Hayashi:2023fgl}.
The UV renormalons, however, exist in $\tilde{C}_n(\tau)$,
which do not induce inevitable uncertainties.

On the other hand, one can derive 
\be
\tilde{\rho}(\tau)=\sum_{n=1}^{\infty} \frac{c_n}{n!} \lt(\frac{\tau}{m^2} \rt)^n
\ee
with $c_n$ of eq.~\eqref{cnrho}.
This can be considered an exact expression of $\tilde{\rho}(\tau)$.
The reasoning behind these conclusions is parallel to the first example; see the paragraph of eq.~\eqref{tildeDalphalow}.

\subsection{Duality violations}
\label{sec:3.3}

DV are found also in this example and their origin can be understood in a manner almost parallel to the first example.
However, we note a minor difference from the previous example:
while the contribution from the small arc around $r=-1/m_{\rm gap}^2$ in the integration contour of the bottom-right panel 
can be neglected in the first example, this is not the case in the present example.

The exact result of $\tilde{\rho}(\tau)$ based on the integral along the bottom-right panel of fig.~\ref{contours}
is given by\fn{
A careful treatment of the contribution from the small arc around $r=0$ 
is required as noted in footnote~\ref{fn2}.}
\be
\tilde{\rho}(\tau)
=\frac{1}{\pi} \int_{0+i0}^{-1/(9 m^2)+\epsilon+i0} \frac{dr}{r} {\rm Im} [\rho(1/r)] e^{\tau r}
+\frac{1}{2 \pi i} \int_{C} \frac{dr}{r} \rho(1/r) e^{\tau r} \label{tilderhoexactanother}
\ee
where $C$ is the small circle around $r=-1/(9m^2)$, on which $r=\epsilon e^{i \theta}-1/(9m^2)$ with $\epsilon \ll1$ and  $0 < \theta < 2 \pi$.
If $\Pi_{\sigma}(q^2)$ is assumed to behave around the singularity as
\be 
\Pi_{\sigma}(q^2) \simeq A [(q^2+9m^2)+\mathcal{O}(q^2+9m^2)^2] \log{(q^2+9m^2)} \label{assumption}
\ee
with a constant $A$,
we obtain
\begin{align}
\frac{1}{2 \pi i} \int_{C} \frac{dr}{r} \rho(1/r) e^{\tau r}
&=-Ae^{-\tau/(9m^2)} \label{intC}
\end{align}
from the most singular term $\rho(q^2) \simeq A \frac{q^2}{q^2+9m^2}$.  
Therefore, using 
\be
\tilde{\rho}_{\rm OPE}(\tau)=\frac{1}{\pi} \int_{0+i0}^{-\infty+i0} \frac{dr}{r} \, {\rm Im} [\rho_{\rm OPE}(1/r)] e^{\tau r}
\ee
and the fact that $\rho(1/r)=\rho_{\rm OPE}(1/r)$ for $-1/(9m^2) < r < 0$,
we obtain a relation
\be
\tilde{\rho}(\tau)=\tilde{\rho}_{\rm OPE}(\tau)+\tilde{\rho}_{\rm DV}(\tau) ,
\ee
with
\be
\tilde{\rho}_{\rm DV}(\tau)=-\frac{1}{\pi} \int_{-1/(9m^2)+i0}^{-\infty+i0} \frac{dr}{r} {\rm Im} [\rho_{\rm OPE}(1/r)] e^{\tau r}-A  e^{-\tau/(9m^2)} . \label{tilderhoDV}
\ee
A discussion to support the assumption~\eqref{assumption} is given in App.~\ref{app:C}.

The approximate function for the first term of eq.~\eqref{tilderhoDV} can be obtained in a manner parallel 
to the previous example using the Borel transform:
\begin{align}
B_{\tilde{\rho}_{\rm DV}}(u)
&\equiv -\frac{1}{\pi} \int_{-1/(9m^2)+i0}^{-\infty+i0} \frac{dr}{r} {\rm Im} \, [ B_{\rho_{\rm OPE}}(u)|_{q^2 \to 1/r}] e^{\tau r}  \non
&=-\frac{1}{4 \pi^2} \lt(\frac{\mu^2}{\tau} \rt)^u \sum_{n=0}^{\infty} \lt(\frac{m^2}{\tau} \rt)^n \non
&\quad{} \times \bigg\{ \lt[c_n(u) (\Gamma(u+n,\tau/(9m^2) \log{(\tau/m^2)}-\del_u \Gamma(u+n, \tau/(9m^2))+d_n^1(u) \Gamma(u+n,\tau/(9m^2))\rt] \non
&\qquad{} \times \sin{[\pi (u+n)]}  \non
&\qquad{}-(c_n(u)+d_n^2(u)) \pi \cos{[\pi (u+n)] \Gamma(u+n,\tau/(9m^2))} \bigg\} . \label{tilderhoDVBorel}
\end{align}
One can see that $\log{(\tau/m^2)}$ terms do not appear in the approximate function, repeating a calculation similar to eq.~\eqref{tildeDalphaDVapprox}.
Overall, the approximate function of $\tilde{\rho}_{\rm DV}(\tau)$ is given by 
\be
\tilde{\rho}_{\rm DV}(\tau)=K e^{-\tau/(9m^2)} \lt[1+k_1 \lt( \frac{9m^2}{\tau} \rt)+k_2 \lt( \frac{9m^2}{\tau} \rt)^2+\cdots \rt] , \label{tilderhoDVapprox}
\ee
where the function form is almost kept parallel to the first example (see eq.~\eqref{tildeDalphaDVapprox}).

From this calculation, the existence of the extra contribution is attributed to the fact that
the derivatives of $\Pi_{\sigma}(q^2)$ is considered rather than $\Pi_{\sigma}(q^2)$ itself.
The imaginary part of the latter rises from zero at the singular point
and the contribution around $r=-1/m_{\rm gap}^2$ can be neglected.

\subsection{Low energy limit from the OPE}

We extract the first non-vanishing coefficient in the low energy expansion, $c_1=-0.012396...$,
with reference to the formula
\be
c_1 =\frac{m^2}{2 \pi i} \oint_{\tau=\tau_0 e^{i \theta}} \frac{d \tau}{\tau^2} \lt(\frac{\tau}{\tau_0}+1 \rt)^k \tilde{\rho}(\tau) .
\ee
Here the integral contour surrounds the origin ($\tau_0>0$) and $k$ is a positive integer $k$.
Employing the OPE, we extract $c_1$ by
\be
c_1 \approx \frac{m^2}{2 \pi i} \oint_{\tau=\tau_0 e^{i \theta}} \frac{d \tau}{\tau^2} \lt(\frac{\tau}{\tau_0}+1 \rt)^k \tilde{\rho}_{\rm OPE}(\tau)  \label{c1est} .
\ee
From eq.~\eqref{tilderhoDVapprox} and calculations similar to eq.~\eqref{suppressDV},
one sees that the optimal choice of $k$ is $k_{\rm opt}=\tau_0/(9m^2)$.

\begin{figure}[tb]
\begin{minipage}{0.5\hsize}
\begin{center}
\includegraphics[width=7cm]{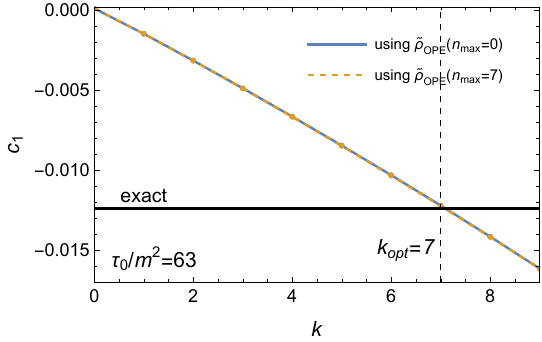}
\end{center}
\end{minipage}
\begin{minipage}{0.5\hsize}
\begin{center}
\includegraphics[width=7cm]{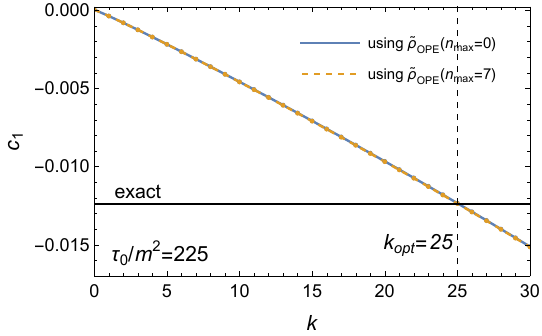}
\end{center}
\end{minipage}
\caption{Extraction of $c_1$ based on eq.~\eqref{c1est} using $\tilde{\rho}_{\rm OPE}(\tau)$. 
The radius of the integral contour is taken as
$\tau_0/m^2=63$ (left) and $\tau_0/m^2=225$ (right), respectively. 
The horizontal axis shows $k$ in eq.~\eqref{c1est} and
the point corresponding to the optimal choice $k_{\rm opt}=\tau_0/(9m^2)$ is shown in each case.
The truncation orders of the OPE (denoted by $n_{\rm max}$) are indicated in the figures. 
The exact answer $c_1=-0.012396...$ is shown by the black lines.}
\label{figc1rho}
\end{figure}

The results of $c_1$ extracted based on eq.~\eqref{c1est} are shown 
in fig.~\ref{figc1rho}.
We take the radius of the integral contour as $\tau_0/m^2=63$ (left) and $\tau_0/m^2=225$ (right),
where the optimal $k$ is given by $k_{\rm opt}=7$ and $k_{\rm opt}=25$, respectively.
In the figure, the results with different OPE truncation orders are shown.
One can see that the truncated OPE with $n_{\rm max}=0$ already gives 
an estimate with precision about 1.5 \% (left) and 0.5 \% (right) with the optimal choices of $k$.
The higher order results almost overlap with the $n_{\rm max}=0$ result.

\begin{figure}[tb]
\begin{center}
\includegraphics[width=8cm]{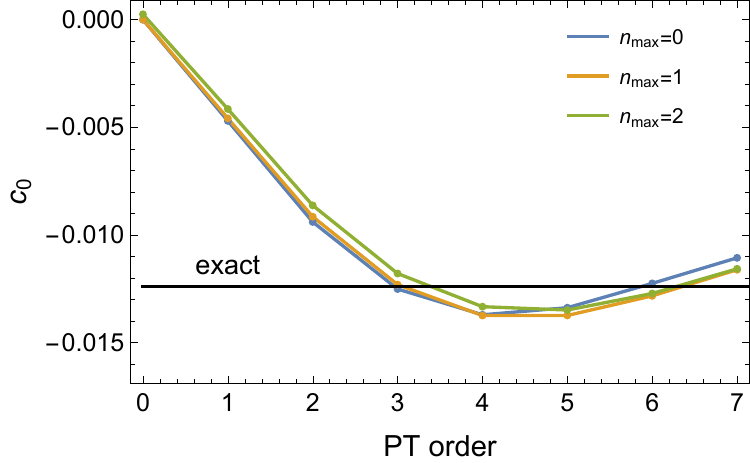}
\end{center}
\caption{Extraction of $c_1$ based on eq.~\eqref{c1est} with $\tau_0/m^2=1350$, using 
fixed order perturbation theory for the Wilson coefficients.
The blue line is the estimate where the OPE is truncated at $n_{\rm max}=0$
and the Wilson coefficient $\tilde{C}_0(\tau)$ is calculated by fixed order perturbation theory;
the order of the perturbation theory is indicated by ``PT order''. 
(``PT order''$=i$ corresponds to truncating the series at $\mathcal{O}(\hat{g}^{i+1})$.)
The orange and green lines are obtained for the OPE truncation order at $n_{\rm max}=1,2$, respectively.
For all the relevant Wilson coefficients $\tilde{C}_n(\tau)$, the perturbative series are truncated 
at the same order indicated by ``PT order''.
$k$ in eq.~\eqref{c0withk} is fixed to the optimal value $k=\tau_0/(9m^2)=150$.
The exact answer $c_0=-0.012396...$ is shown by the black lines.}
\label{figc1rho1350PT}
\end{figure}

We also show a result 
using fixed order perturbation theory for the Wilson coefficients in $\tilde{\rho}_{{\rm OPE}}(\tau)$, 
rather than using the Borel integral.
It is notable that the fact that the Wilson coefficients are free from IR renormalons
allows one to employ fixed order perturbation theory, which is a practical calculation method, 
as a more systematic tool, although one needs to pay attention to UV renormalons.
In fig.~\ref{figc1rho1350PT}, we show the result with $\tau_0/m^2=1350$,
fixing $k$ to its optimal value $k=1350/9=150$.
The horizontal axis represents the truncation order of the perturbative series for the Wilson coefficients $\tilde{C}_n(\tau)$,
while the truncation order of the OPE is fixed for each colored line.
We take the renormalization scale $\mu$ as $\mu^2=\tau_0=1350 m^2$
for the perturbative series.
Already at the OPE truncation order $n_{\rm max}=0$ and ``PT order''$=2$, 
corresponding to the perturbative series of $\tilde{C}_0(\tau)$ truncated at $\mathcal{O}(\hat{g}^3)$,
the error of the result is only 25 \%.
The higher orders in the OPE give minor corrections, which is consistent with the observation in fig.~\ref{figc1rho}.

\section{Conclusions and discussion}

In this paper, we proposed a method to extract the low energy limit of physical observables 
from their high energy expansions, calculable via the OPE.
In this method, we apply the inverse Laplace transform to a observable $D(p^2)$ 
with respect to the inverse of the typical scale, i.e., $1/p^2$, and obtain
the transformed quantity, denoted by $\tilde{D}(\tau)$.
The remarkable feature is that  $\tilde{D}(\tau)$ is analytic at any point in the $\tau$-plane, as long as 
the original observable $D(p^2)$ is analytic for $|p^2|<m_{\rm gap}^2$ due to a non-zero mass gap.
Combining this property and the fact that the small-$\tau$ expansion of $\tilde{D}(\tau)$ has a simple relation with
the low energy (small-$p^2$) expansion of $D(p^2)$,
the low energy expansion coefficients become accessible from the high energy expansion of  $\tilde{D}(\tau)$.
We particularly employed the residue theorem to extract the low energy limit of $D(p^2)$,
using the OPE result of $\tilde{D}(\tau)$ for the integrand of the contour integral.
We examined this idea using the two-dimensional $\mathcal{O}(N)$ nonlinear $\sigma$ model, 
which is solvable at the large-$N$ limit and thus serves as a good testing ground.

Through the study of the two correlators in the model,  it was revealed that the OPE for $\tilde{D}(\tau)$ possesses the following features:
(i) $\tilde{D}_{\rm OPE}(\tau)$ is a divergent asymptotic series expansion in $m^2/\tau$, where $m^2$ is the dynamical scale of the model,
and (ii) it requires the correction called DV, characterized by the approximate function form $e^{-\tau/m_{\rm gap}^2}/\tau^{i}$ with an integer $i$.
Considering the second feature, we used the equation like eq.~\eqref{c0modified}
to properly suppress DV in estimate of the low energy limit.
It was argued that the optimal choice of $k$ in eq.~\eqref{c0modified} is given by $k_{\rm opt}=\tau_0/m_{\rm gap}^2$,
where $\tau_0$ is the radius of the integration contour.
Once $k_{\rm opt}$, or equivalently $m_{\rm gap}$, is assumed to be known, 
the low energy limits can be accurately extracted; see figs.~\ref{figc0Dalpha},\ref{figc0Dalpha600PT},\ref{figc1rho},\ref{figc1rho1350PT}.
In particular, figs.~\ref{figc0Dalpha600PT} and \ref{figc1rho1350PT} are encouraging from the viewpoint of practicality;
they show that the estimates of the low energy limits are reasonable already at relatively small orders of perturbation theory and the OPE.
It was also pointed out that $m_{\rm gap}$ or then $k_{\rm opt}$ can be detected by the asymptotic divergent behavior of $\tilde{D}_{\rm OPE}(\tau)$,
without relying on the nonperturbative answers.

One of the remarkable features of the inverse Laplace transform is that 
this transform gives the OPE which is free from the renormalon problem, 
as clarified in ref.~\cite{Hayashi:2023fgl}.
In the second example studied in this paper, 
the OPE for the original quantity $\rho(q^2)$ possesses renormalon uncertainties, 
while $\tilde{\rho}(\tau)$, obtained after the inverse Laplace transform, 
does not possess IR renormalons at all.
This feature renders fixed order perturbation theory a more systematic approximation method for $\tilde{D}_{\rm OPE}(\tau)$
and also obviates the need to determine ambiguous matrix elements using nonperturbative data or calculations;
the ambiguous matrix elements $\langle \mathcal{O}^{\delta \alpha}_{n, j} \rangle$ in eq.~\eqref{rhoOPE}
indeed disappears after the inverse Laplace transform (see eq.~\eqref{tilderhoOPE}). 
Therefore, the OPE calculation after the inverse Laplace transform proceeds much more straightforwardly.
Indeed, it was demonstrated that fixed order perturbation theory allows us to estimate the low energy limit accurately.

We briefly discuss a possible issue when the method is applied to QCD.
In the large-$N$ limit of the $O(N)$ nonlinear $\sigma$ model,
 the convergence radius of the OPE is finite, while that of the OPE 
 after the inverse Laplace transform becomes zero, which then requires the correction called DV.
However, in QCD, the OPE before the inverse Laplace transform is already divergent and
requires DV corrections  \cite{Shifman:2000jv}.
Thus, we need to reconsider the discussions on DV, developed
based on the convergence of the OPE before the inverse Laplace transform.
This issue needs to be addressed for the application of the method to QCD.

\acknowledgments
 
The author thanks Yuuki Hayashi, Go Mishima, and Yukinari Sumino as the present work is inspired by 
the work~\cite{Hayashi:2023fgl} in collaboration with them.
The author is the Yukawa Research Fellow supported by Yukawa Memorial Foundation.
This work is also supported by JSPS Grant-in-Aid for Scientific Research Grant Numbers, 
JP19K14711 and JP23K13110.

\appendix

\section{Borel transform and Borel integral}
\label{app:A}

\subsection{Definition and review}

Given a perturbative series, 
\be
\sum_{n=0}^{\infty} a_{n} \hat{g}^{n+1}(\mu^2) , \label{pertseries}
\ee
we define the Borel transform by
\be
B(u) \equiv \sum_{n=0}^{\infty} \frac{a_n}{n!} u^n .
\ee
This is the generating function of the perturbative coefficients 
\be
a_n=\frac{d^n}{d u^n} B(u) \bigg|_{u \to 0} .
\ee
The all-order sum of the perturbative series~\eqref{pertseries} is given by the Borel integral
\be
\int_0^{\infty} du \, B(u) e^{-u/\hat{g}(\mu^2)} . \label{Borelint}
\ee

When the perturbative series~\eqref{pertseries} has renormalon divergences
\be
a_n \sim N_1 v_1^{-n} n!+N_2 v_2^{-n} n!+ \cdots \quad{} {\text{for $n \gg 1$}} \label{asym}
\ee
where $v_1$, $v_2$,... are real parameters and $N_1$, $N_2$ ... are constants, 
one sees that the Borel transform has poles at $u=v_1$, $v_2$,...
The singular points of the Borel transform thus correspond to the parameters $v_i$ in eq.~\eqref{asym}
and the smallest $|v_i|$ determines the leading behavior of the perturbative coefficients.
If the Borel transform does not have any singularities,
there are no renormalon divergences in the original perturbative series.
The positive singular points are called IR renormalons, corresponding to sign-non-alternating divergent behaviors,
while the negative singular points are called UV renormalons, corresponding to sign-alternating divergent behaviors.
If IR renormalons exist the Borel integral~\eqref{Borelint} needs to be regularized, because the singular points appear on the integration path,
and it has ambiguities depending on how the contour is deformed.
However, this case is not very relevant to the present paper, because IR renormalons disappear after the inverse Laplace transform.
UV renormalons do not induce ambiguities to the Borel integral~\eqref{Borelint}.

\subsection{Regularization of Borel integral}

We need to regularize the Borel integral which does not converge due to a divergent behavior at $u \to \infty$.
In this paper, the Borel integral of interest has form of
\be
\int_0^{\infty} du \, f(u) \Gamma(u+n_1,z) \sin{[\pi (u+n_2)]} e^{-u \log{(\tau/m^2)}} ,
\ee
where $n_1$ and $n_2$ are (integer) constants. $z=\tau/m_{\rm gap}^2$ or $z=0$,
 where the incomplete gamma function reduces to the gamma function. 
We can assume that $f(u) \Gamma(u+n_1,z) \sin{[\pi (u+n_2)]}$ does not have singularities at $u>0$
 for all the examples studied in this paper.
We can rewrite this integral as
\begin{align}
&\int_0^{\infty} du \, f(u) \Gamma(u+n_1,z) \sin{[\pi (u+n_2)]} e^{-u \log{(\tau/m^2)}}   \non
&=\lim_{\epsilon \to 0} \lt[\int_{\epsilon}^{u_{\rm sing}-\epsilon}+\int_{u_{\rm sing}+\epsilon}^{\infty} \rt] 
du\,  \frac{1}{2i} f(u) \Gamma(u+n_1,z) e^{i \pi (u+n_2)} e^{-u \log{(\tau/m^2)}}  \non
&\quad{}
-\lim_{\epsilon \to 0} \lt[\int_{\epsilon}^{u_{\rm sing}-\epsilon}+\int_{u_{\rm sing}+\epsilon}^{\infty} \rt] 
du\, \frac{1}{2i} f(u) \Gamma(u+n_1,z) e^{-i \pi (u+n_2)} e^{-u \log{(\tau/m^2)}}  \non 
&=\lim_{\epsilon \to 0} \int_{\epsilon}^{\infty+i0} 
du\,  \frac{1}{2i} f(u) \Gamma(u+n_1,z) e^{i \pi (u+n_2)} e^{-u \log{(\tau/m^2)}}  \non
&\quad{}-\lim_{\epsilon \to 0} \int_{\epsilon}^{\infty-i0} 
du\,  \frac{1}{2i} f(u) \Gamma(u+n_1,z) e^{-i \pi (u+n_2)} e^{-u \log{(\tau/m^2)}}   \non
&\quad{} +\frac{\pi}{2} {\rm Res}[f(u) \Gamma(u+n_1,z) e^{i \pi (u+n_2)} e^{-u \log{(\tau/m^2)}} ] |_{u=u_{\rm sing}} \non
&\quad{}
+\frac{\pi}{2} {\rm Res}[f(u) \Gamma(u+n_1,z) e^{-i \pi (u+n_2)} e^{-u \log{(\tau/m^2)}} ] |_{u=u_{\rm sing}} .
\end{align}
Note that the functions $f(u) \Gamma(u+n_1,z) e^{\pm i \pi (u+n)}$ can have poles on the integration path
because $e^{\pm i \pi (u+n)}$ does not have zeros unlike $ \sin[{\pi (u+n)}]$,
and we assume that they are located at $u=0$ and $u=u_{\rm sing}$.
(The formulae presented here are valid if these singularities are absent.)
The integration contour $\epsilon \to \infty \pm i0$ represents the contour that continuously connects $0$ to $\infty$,
lying slight above or below the real axis.
The third and forth terms subtract the extra contributions arising from the change of the integration paths.
We regularize the integrals as
\begin{align}
&\lim_{\epsilon \to 0}  \int_{\epsilon}^{\infty+i0} du \, \frac{1}{2i}f(u) \Gamma(u+n_1,z) e^{ i \pi (u+n_2)} e^{-u \log{(\tau/m^2)}}  \non
&\overset{\rm reg.}{=}
\lim_{\epsilon \to 0}  \int_{\epsilon}^{\infty} dt \, \frac{1}{2} f(it) \Gamma(it+n_1,z) e^{- \pi t} e^{i \pi n_2} e^{-i t \log{(\tau/m^2)}} \non
&\quad{}+\frac{\pi}{4} {\rm Res} \,[f(u) \Gamma(u+n_1,z) e^{ i \pi (u+n_2)} e^{-u \log{(\tau/m^2)}} ]|_{u=0}  ,
\end{align}
and
\begin{align}
&-\lim_{\epsilon \to 0}  \int_{\epsilon}^{\infty-i0} du \, \frac{1}{2i}f(u) \Gamma(u+n_1,z) e^{-i \pi (u+n_2)}  e^{-u \log{(\tau/m^2)}}  \non
&\overset{\rm reg.}{=}
\lim_{\epsilon \to 0}  \int_{\epsilon}^{\infty} dt \, \frac{1}{2} f(-it) \Gamma(-it+n_1,z) e^{- \pi t} e^{-i \pi n_2} e^{it \log{(\tau/m^2)}}  \non
&\quad{}+\frac{\pi}{4} {\rm Res} \,[f(u) \Gamma(u+n_1,z) e^{- i \pi (u+n_2)} e^{-u \log{(\tau/m^2)}} ]|_{u=0},
\end{align}
rotating the integration contour to the positive or negative imaginary axis.
These regularizations correspond to neglecting the contribution from the large arc $u=R e^{i \theta}$ with $0 \leq \theta \leq \pi/2$ or $-\pi/2 \leq \theta \leq 0$ ($R \gg 1$).
One can show that the $t$ integrals converge for any $\tau=|\tau| e^{i \theta}$ with $-\pi \leq \theta \leq \pi$
for $f(u)$ considered in this paper.
The second terms are contribution from the small arc $u=\epsilon e^{i \theta}$, arising upon rotating the contours.
To summarize, the regularized Borel integral is given by
\begin{align}
&\int_0^{\infty} du \, f(u) \Gamma(u+n_1,z) \sin{[\pi (u+n_2)]} e^{-u \log{(\tau/m^2)}} \non
&\overset{\rm reg.}{=}
\lim_{\epsilon \to 0}  \int_{\epsilon}^{\infty} dt \, \frac{1}{2} f(it) \Gamma(it+n_1,z) e^{- \pi t} e^{i \pi n_2} e^{-it \log{(\tau/m^2)}} \non 
&\quad{}+\lim_{\epsilon \to 0}  \int_{\epsilon}^{\infty} dt \, \frac{1}{2} f(-it) \Gamma(-it+n_1,z) e^{- \pi t} e^{-i \pi n_2} e^{it \log{(\tau/m^2)}}  \non
&\quad{}
+\frac{\pi}{4} {\rm Res}  \,[f(u) \Gamma(u+n_1,z) (e^{ i \pi (u+n_2)}+e^{- i \pi (u+n_2)}) e^{-u \log{(\tau/m^2)}} ] \big|_{u=0} \non
&\quad{}
+\frac{\pi}{2} {\rm Res}[f(u) \Gamma(u+n_1,z) (e^{i \pi (u+n_2)}+e^{-i \pi (u+n_2)}) e^{-u \log{(\tau/m^2)}} ] \big|_{u=u_{\rm sing}} . \label{Boreintregularize}
\end{align}

In calculating the regularized Borel integral of $\tilde{C}_0(\tau)$ for $\tilde{D}_{\alpha}(\tau)$,
the $u_{\rm sing}=1$ singularity appears,
while the singularitiy at $u=0$ appears in calculating the regularized Borel integrals for 
$\tilde{C}_0(\tau)$ and $\tilde{C}_1(\tau)$ in $\tilde{D}_{\alpha, {\rm OPE}}(\tau)$.

In the second example studied in sec.~\ref{sec:3}, a cosine function appears (eqs.~\eqref{tilderhoBorel} and \eqref{tilderhoDVBorel}). 
In this case, rewriting $\cos{[\pi (u+n_2)]}=(e^{i \pi (u+n_2)}+e^{-i \pi (u+n_2)})/2$ and performing a similar contour deformation, 
we can regularize it by
\begin{align}
&\int_0^{\infty} du \, g(u) \Gamma(u+n_1,z) \cos{[\pi (u+n_2)]} e^{-u \log{(\tau/m^2)}} \non
&\overset{\rm reg.}{=}{\lim}_{\epsilon \to 0} \int_{\epsilon}^{\infty} dt \, \frac{i}{2} g(it) \Gamma(it+n_1,z) e^{-\pi t} e^{i \pi n_2} e^{-it \log{(\tau//m^2)}} \non
&\quad{}-{\lim}_{\epsilon \to 0} \int_{\epsilon}^{\infty} dt \, \frac{i}{2} g(-it) \Gamma(-it+n_1,z) e^{-\pi t} e^{-i \pi n_2} e^{it \log{(\tau//m^2)}}  \label{Boreintregularize2}
\end{align}
In this example, the singularity at $u=0$ or $u_{\rm sing}>0$ does not appear.

Actually the above regularized Borel integral can be considered as a part of a well-defined (convergent) Borel integral.
According to the discussion in sec.~\ref{sec:2.3}, the quantity which converges to the exact answer is given by the form of
\be
\int_0^{\infty} du \, f(u) \gamma(u+n_1,\tau/m_{\rm gap}^2) \sin{[\pi (u+n_2)]} e^{-u \log{(\tau/m^2)}}
\ee
from eq.~\eqref{OPEplusDV} and eqs.~\eqref{tildeDalphaBorel} and \eqref{tildeDalphaDVBorel}.
Note that $\gamma(u+n_1,z)=\Gamma(u+n_1)-\Gamma(u+n_1,z)$.
The asymptotic behavior $\gamma(u+n_1,z) \sim z^{u}$ at large $u$ indicates that 
the above $u$-integral converges as long as $m_{\rm gap}/m >1$.
Also it is legitimate to deform the contours as above after splitting the sine function:
\begin{align}
&\int_0^{\infty} du \, f(u) \gamma(u+n_1,\tau/m_{\rm gap}^2) \sin{[\pi (u+n_2)]} e^{-u \log{(\tau/m^2)}} \non
&=
\lim_{\epsilon \to 0}  \int_{\epsilon}^{\infty} dt \, \frac{1}{2} f(it) \gamma(it+n_1,\tau/m_{\rm gap}^2) e^{- \pi t} e^{i \pi n_2} e^{-it \log{(\tau/m^2)}} \non 
&\quad{}+\lim_{\epsilon \to 0}  \int_{\epsilon}^{\infty} dt \, \frac{1}{2} f(-it) \gamma(-it+n_1,\tau/m_{\rm gap}^2) e^{- \pi t} e^{-i \pi n_2} e^{it \log{(\tau/m^2)}}  \non
&\quad{}
+\frac{\pi}{4} {\rm Res}  \,[f(u) \gamma(u+n_1,\tau/m_{\rm gap}^2) (e^{ i \pi (u+n_2)}+e^{- i \pi (u+n_2)}) e^{-u \log{(\tau/m^2)}} ] \big|_{u=0} \non
&\quad{}
+\frac{\pi}{2} {\rm Res}[f(u) \gamma(u+n_1,\tau/m_{\rm gap}^2) (e^{i \pi (u+n_2)}+e^{-i \pi (u+n_2)}) e^{-u \log{(\tau/m^2)}} ] \big|_{u=u_{\rm sing}} . 
\end{align}
Substituting $\gamma(u+n_1,z)=\Gamma(u+n_1)-\Gamma(u+n_1,z)$ in this equation, one indeeds find 
the contribution of eq.~\eqref{Boreintregularize} with $z=0$ and that with $z=\tau/m_{\rm gap}^2$.
The regularization of the latter integral, where the integral of $g(u) \Gamma(u+n_1,z) \cos[\pi(u+n_2)] e^{-u \log{(\tau/m^2)}}$ is considered,
can also be understood in a parallel manner.

\section{Formulae of the inverse Laplace transform}
\label{app:B}

To calculate the inverse Laplace transform, the following formula is repeatedly used: 
\be
\frac{1}{2 \pi i} \int_{-i\infty}^{i \infty} \frac{dr}{r^{\kappa}} r^{a} e^{\tau r}
=\tau^{\kappa-1-a} \frac{1}{\Gamma(\kappa-a)} 
=\tau^{\kappa-1-a} \frac{1}{\pi} \Gamma(a-\kappa+1) \sin{[\pi (\kappa-a)]} . \label{inverseLaplaceformula}
\ee
Taking the derivative with respect to $a$ of the above formula leads to
\begin{align}
&\frac{1}{2 \pi i} \int_{-i\infty}^{i \infty} \frac{dr}{r^{\kappa}} r^{a} \log{r} \, e^{\tau r} \non
&=\tau^{\kappa-1-a} \frac{1}{\pi} (-\log{\tau}+\psi(a-\kappa+1)) \Gamma(a-\kappa+1) \sin{[\pi (\kappa-a)]} \non
&\quad{}-\tau^{\kappa-1-a} \Gamma(a-\kappa+1) \cos{[\pi(\kappa-a)]} . \label{inverseLaplaceformula2}
\end{align}

\section{Supplementary analysis regarding eq.~\eqref{assumption}}
\label{app:C}

 To examine the assumption~\eqref{assumption}, we give the exact result of ${\rm Im} \, \Pi_{\sigma}(q^2)$.
To this end, we first perform the $k_0$-integral of eq.~\eqref{Pi}.
For convenience, we use the Minkowski metric only in this appendix and set $q=(q_0,0)$. 
The $k_0$-integral gives three contributions by closing the $k_0$-integral contour below:
one is the pole contribution from the  propagator $1/[m^2-(q+k)^2]$,
another is the cut contribution from $D_{\alpha}$, and the other is the large arc contribution
where $k_0=R e^{i \theta}$ with $R \gg1$ and $-\pi<\theta<0$.
It can be shown that the first contribution gives non-zero and finite imaginary part 
while the second contribution does not give imaginary part.
Although the third contribution is divergent, 
this should not give $q^2$-dependent imaginary part.
Actually the finiteness of the imaginary part of ${\rm Im} \, \Pi_{\sigma}(q^2)$
can be inferred from the result of ref.~\cite{Beneke:1998eq};
the UV divergences are only encoded in the $C_n^2$ terms, 
which never generate imaginary parts due to the absence of the logarithmic terms $\sim \log{(q^2/\mu^2)}$.
In this way, we obtain a one-dimensional ($k_1$) integral formula of ${\rm Im} \, \Pi^M_{\sigma}(q^2)$ for $q^2=q_0^2>9m^2$,
\begin{align}
{\rm Im} \, \Pi^M_{\sigma}(q^2+i0) 
&={\rm Im} \lt[ \int \frac{d k_0}{2 \pi i} \frac{dk_1}{2 \pi} \frac{1}{m^2-(q+k)^2} D^M_{\alpha}(k^2) \rt] \non
&=\int_0^{\frac{1}{2 q_0} \sqrt{(q_0^2-9m^2)(q_0^2-m^2)}} \frac{d k_1}{2 \pi}  \frac{1}{\sqrt{k_1^2+m^2}}  \non
&\qquad{}\times\lt(4 \pi \sqrt{k^2 (k^2-4m^2)} \frac{-\pi}{\log^2{\lt[\frac{\sqrt{k^2}+\sqrt{k^2-4m^2}}{\sqrt{k^2}-\sqrt{k^2-4m^2}} \rt]}+\pi^2}  \Bigg|_{k^2=(-q_0+\sqrt{k_1^2+m^2})^2-k_1^2}\rt) ,
\label{numImPi}
\end{align}
where
\be
D^M_{\alpha}(k^2)=\frac{4 \pi \sqrt{(4m^2-k^2)(-k^2)}}{\log{\lt[\frac{\sqrt{4m^2-k^2}+\sqrt{-k^2}}{\sqrt{4m^2-k^2}-\sqrt{-k^2}} \rt]} } .
\ee
The above $k_1$-integral can be evaluated numerically.
The superscript ``{\it{M}}'' represents that the corresponding quanitities are defined in the Minkwoski metric. 
We numerically confirmed that ${\rm Im} \, \Pi^M_{\sigma}(q^2+i0) $ has a consistent behavior with the assumption~\eqref{assumption};
the imaginary part rises linearly at $q^2=9m^2$.

We also made another observation.
Using integration-by-parts and eq.~\eqref{intC},
we obtain from eq.~\eqref{tilderhoexactanother}
\be
\tilde{\rho}(\tau)
= -\frac{1}{\pi} \int_{9m^2+i0}^{\infty+i0} dq^2 \frac{\tau (\tau-2 q^2)}{(q^2)^2}  {\rm Im} \, [\Pi^M_{\sigma}(q^2)] e^{-\tau/(9m^2)}
+\frac{1}{\pi} {\rm Im} [{\Pi^M_{\sigma}}'(9m^2+i0) ] e^{-\tau/(9m^2)}-A e^{-\tau/(9m^2)} .
\ee
Using the assumption~\eqref{assumption}, we have  ${\rm Im} \, [{\Pi^M_{\sigma}}'(9m^2+i0)) ]=\pi A$
noting that $\Pi^M_{\sigma}(q^2) \approx A(9m^2-q^2) \log{(9m^2-q^2)}$, and therefore
\be
\tilde{\rho}(\tau)
= -\frac{1}{\pi} \int_{9m^2+i0}^{\infty+i0} dq^2 \frac{\tau (\tau-2 q^2)}{(q^2)^2}  {\rm Im} \, [\Pi^M_{\sigma}(q^2)] e^{-\tau/q^2} .
\ee
We numerically confirmed that this indeed reproduces the exact result of $\tilde{\rho}(\tau)$ given by the Taylor series.
These observations seem to support the assumption \eqref{assumption} 
and also the conclusions following from it, as given in Sec.~\ref{sec:3.3}.

\bibliographystyle{utphys}
\bibliography{bib}

\providecommand{\href}[2]{#2}\begingroup\raggedright\begin{thebibliography}{10}

\bibitem{Bardeen:1976zh}
W.~A. Bardeen, B.~W. Lee, and R.~E. Shrock, ``{Phase Transition in the
  Nonlinear $\sigma$ Model in 2 + $\epsilon$ Dimensional Continuum},''
  \href{http://dx.doi.org/10.1103/PhysRevD.14.985}{{\em Phys. Rev. D}
  {\bfseries 14} (1976) 985}.

\bibitem{Hayashi:2023fgl}
Y.~Hayashi, G.~Mishima, Y.~Sumino, and H.~Takaura, ``{Renormalon subtraction in
  OPE by dual space approach: nonlinear sigma model and QCD},''
  \href{http://dx.doi.org/10.1007/JHEP06(2023)042}{{\em JHEP} {\bfseries 06}
  (2023) 042}, \href{http://arxiv.org/abs/2303.16392}{{\ttfamily
  arXiv:2303.16392 [hep-ph]}}.

\bibitem{Beneke:1998ui}
M.~Beneke, ``{Renormalons},''
  \href{http://dx.doi.org/10.1016/S0370-1573(98)00130-6}{{\em Phys. Rept.}
  {\bfseries 317} (1999) 1--142},
  \href{http://arxiv.org/abs/hep-ph/9807443}{{\ttfamily arXiv:hep-ph/9807443}}.

\bibitem{David:1982qv}
F.~David, ``{Nonperturbative Effects and Infrared Renormalons Within the 1/$N$
  Expansion of the O($N$) Nonlinear $\sigma$ Model},''
  \href{http://dx.doi.org/10.1016/0550-3213(82)90266-8}{{\em Nucl. Phys. B}
  {\bfseries 209} (1982) 433--460}.

\bibitem{Beneke:1998eq}
M.~Beneke, V.~M. Braun, and N.~Kivel, ``{The Operator product expansion,
  nonperturbative couplings and the Landau pole: Lessons from the O(N) sigma
  model},'' \href{http://dx.doi.org/10.1016/S0370-2693(98)01339-2}{{\em Phys.
  Lett. B} {\bfseries 443} (1998) 308--316},
  \href{http://arxiv.org/abs/hep-ph/9809287}{{\ttfamily arXiv:hep-ph/9809287}}.

\bibitem{Schubring:2021hrw}
D.~Schubring, C.-H. Sheu, and M.~Shifman, ``{Treating divergent perturbation
  theory: Lessons from exactly solvable 2D models at large N},''
  \href{http://dx.doi.org/10.1103/PhysRevD.104.085016}{{\em Phys. Rev. D}
  {\bfseries 104} no.~8, (2021) 085016},
  \href{http://arxiv.org/abs/2107.11017}{{\ttfamily arXiv:2107.11017
  [hep-th]}}.

\bibitem{Ayala:2019uaw}
C.~Ayala, X.~Lobregat, and A.~Pineda, ``{Superasymptotic and hyperasymptotic
  approximation to the operator product expansion},''
  \href{http://dx.doi.org/10.1103/PhysRevD.99.074019}{{\em Phys. Rev. D}
  {\bfseries 99} no.~7, (2019) 074019},
  \href{http://arxiv.org/abs/1902.07736}{{\ttfamily arXiv:1902.07736
  [hep-th]}}.

\bibitem{Benitez-Rathgeb:2022yqb}
M.~A. Benitez-Rathgeb, D.~Boito, A.~H. Hoang, and M.~Jamin, ``{Reconciling the
  contour-improved and fixed-order approaches for \ensuremath{\tau} hadronic
  spectral moments. Part I. Renormalon-free gluon condensate scheme},''
  \href{http://dx.doi.org/10.1007/JHEP07(2022)016}{{\em JHEP} {\bfseries 07}
  (2022) 016}, \href{http://arxiv.org/abs/2202.10957}{{\ttfamily
  arXiv:2202.10957 [hep-ph]}}.

\bibitem{Beneke:2023wkq}
M.~Beneke and H.~Takaura, ``{Gradient-flow renormalon subtraction and the
  hadronic tau decay series},''
  \href{http://dx.doi.org/10.22323/1.432.0062}{{\em PoS} {\bfseries RADCOR2023}
  (2024) 062}, \href{http://arxiv.org/abs/2309.10853}{{\ttfamily
  arXiv:2309.10853 [hep-ph]}}.

\bibitem{Shifman:1978bx}
M.~A. Shifman, A.~I. Vainshtein, and V.~I. Zakharov, ``{QCD and Resonance
  Physics. Theoretical Foundations},''
  \href{http://dx.doi.org/10.1016/0550-3213(79)90022-1}{{\em Nucl. Phys. B}
  {\bfseries 147} (1979) 385--447}.

\bibitem{Novikov:1984ac}
V.~A. Novikov, M.~A. Shifman, A.~I. Vainshtein, and V.~I. Zakharov,
  ``{Two-Dimensional Sigma Models: Modeling Nonperturbative Effects of Quantum
  Chromodynamics},'' \href{http://dx.doi.org/10.1016/0370-1573(84)90021-8}{{\em
  Phys. Rept.} {\bfseries 116} (1984) 103}.

\bibitem{Shifman:2000jv}
M.~A. Shifman, \href{http://dx.doi.org/10.1142/9789812810458_0032}{``{Quark
  hadron duality},''} in {\em {8th International Symposium on Heavy Flavor
  Physics}}, vol.~3, pp.~1447--1494.
\newblock World Scientific, Singapore, 7, 2000.
\newblock \href{http://arxiv.org/abs/hep-ph/0009131}{{\ttfamily
  arXiv:hep-ph/0009131}}.

\end{thebibliography}\endgroup

\end{document}